\PassOptionsToPackage{unicode}{hyperref}
\PassOptionsToPackage{hyphens}{url}
\documentclass[
]{article}
\usepackage{bm}
\usepackage{mathtools,amssymb}
\mathtoolsset{showonlyrefs}
\usepackage{iftex}
\ifPDFTeX
  \usepackage[T1]{fontenc}
  \usepackage[utf8]{inputenc}
  \usepackage{textcomp} 
\else 
  \usepackage{unicode-math} 
  \defaultfontfeatures{Scale=MatchLowercase}
  \defaultfontfeatures[\rmfamily]{Ligatures=TeX,Scale=1}
\fi
\usepackage{lmodern}
\ifPDFTeX\else
\fi
\IfFileExists{upquote.sty}{\usepackage{upquote}}{}
\IfFileExists{microtype.sty}{
  \usepackage[]{microtype}
  \UseMicrotypeSet[protrusion]{basicmath} 
}{}
\usepackage{xcolor}
\usepackage[margin=1in]{geometry}
\usepackage{longtable,booktabs,array}
\usepackage{calc} 
\usepackage{etoolbox}
\makeatletter
\patchcmd\longtable{\par}{\if@noskipsec\mbox{}\fi\par}{}{}
\makeatother
\IfFileExists{footnotehyper.sty}{\usepackage{footnotehyper}}{\usepackage{footnote}}
\makesavenoteenv{longtable}
\usepackage{graphicx}
\makeatletter
\def\maxwidth{\ifdim\Gin@nat@width>\linewidth\linewidth\else\Gin@nat@width\fi}
\def\maxheight{\ifdim\Gin@nat@height>\textheight\textheight\else\Gin@nat@height\fi}
\makeatother
\setkeys{Gin}{width=\maxwidth,height=\maxheight,keepaspectratio}
\makeatletter
\def\fps@figure{htbp}
\makeatother
\usepackage{setspace}
\usepackage{float}
\usepackage[english]{babel}
\usepackage{cancel}
\usepackage{booktabs}
\usepackage{longtable}
\usepackage{array}
\usepackage{multirow}
\usepackage{wrapfig}
\usepackage{colortbl}
\usepackage{pdflscape}
\usepackage{tabu}
\usepackage{threeparttable}
\usepackage{threeparttablex}
\usepackage[normalem]{ulem}
\usepackage{makecell}
\usepackage{xcolor}

\ifLuaTeX
  \usepackage{selnolig}  
\fi
\usepackage[]{natbib}
\IfFileExists{bookmark.sty}{\usepackage{bookmark}}{\usepackage{hyperref}}
\IfFileExists{xurl.sty}{\usepackage{xurl}}{} 
\hypersetup{
  pdftitle={Active Learning for Simulator Calibration},
  pdfauthor={Scott Koermer, Justin Loda, Aaron Noble, and Robert B. Gramacy},
  hidelinks,
  pdfcreator={LaTeX via pandoc}}

\usepackage{lmodern}        
\usepackage{hyperref}       
\usepackage{url}            

\usepackage{setspace}

\title{
\vspace{-1cm}
Augmenting a simulation campaign\\for hybrid computer model
and field data experiments}
\author{
    Scott Koermer
   \\
    Statistical Sciences Group \\
    Los Alamos National Laboratory \\
  Los Alamos, NM \\
  \texttt{\href{mailto:skoermer@lanl.gov}{\nolinkurl{skoermer@lanl.gov}}} \\
   \and
    Justin Loda
   \\
    Department of Statistics \\
    Virginia Tech \\
  Blacksburg, VA \\
  \texttt{\href{mailto:jbloda@vt.edu}{\nolinkurl{jbloda@vt.edu}}} \\
   \and
    Aaron Noble
   \\
    Department of Mining and Minerals Engineering \\
    Virginia Tech \\
  Blacksburg, VA \\
  \texttt{\href{mailto:aaron.noble@vt.edu}{\nolinkurl{aaron.noble@vt.edu}}} \\
   \and
    Robert B.~Gramacy
   \\
    Department of Statistics \\
    Virginia Tech \\
  Blacksburg, VA \\
  \texttt{\href{mailto:rbg@vt.edu}{\nolinkurl{rbg@vt.edu}}} \\
  }
\date{April 30, 2024}

\begin{document}
\maketitle

\begin{abstract}

The Kennedy and O'Hagan (KOH) calibration framework uses coupled Gaussian
processes (GPs) to meta-model an expensive simulator (first GP), tune its
``knobs'' (calibration inputs) to best match observations from a real
physical/field experiment and correct for any modeling bias (second GP) when
predicting under new field conditions (design inputs). There are
well-established methods for placement of design inputs for data-efficient
planning of a simulation campaign in isolation, i.e., without field data:
space-filling, or via criterion like minimum integrated mean-squared
prediction error (IMSPE). Analogues within the coupled GP KOH framework are
mostly absent from the literature. Here we derive a closed form IMSPE
criterion for sequentially acquiring new simulator data for KOH.  We
illustrate how acquisitions space-fill in design space, but concentrate in
calibration space. Closed form IMSPE precipitates a closed-form gradient for
efficient numerical optimization.  We demonstrate that our
KOH-IMSPE strategy leads to a more efficient simulation campaign on benchmark
problems, and conclude with a showcase on an application to 
equilibrium concentrations of rare earth elements for a liquid-liquid
extraction reaction.

\end{abstract}


\hypertarget{introduction}{%
\section{Introduction}\label{introduction}}

Getting enough real world data to build useful
predictive models is not always possible.  One modern example comes from epidemiology, e.g., Covid-19.    Observational data can be collected when a real epidemic happens; however, ethics reasonably prohibit controlled experimentation --
any at all, let alone ``optimally'' designed.
In many industrial settings, including the chemical processing case motivating this work, collecting real data using an experimental design may be unjustifiably expensive due to production losses, the expense of changing some independent variables, or unsafe working conditions coinciding with the requirements of the procedure. Both of these examples have common thread; a situation where conducting one, or many, additional physical experiments is infeasible due to
prohibitive costs, monetary or otherwise.  
Any analysis based on a small data set of
previous observations -- that may not be representative of the actual situations
or scenarios of interest -- may be severely limited unless some further intervention is taken.

One remedy is to augment the
real-world data with a virtualization of the system
\citep[KOH,][]{kennedy2001bayesian}. (It's acceptable to
infect virtual people with Covid.)  So-called computer simulation
experiments, when calibrated to real world observations, can assist in the
understanding of complex systems that are difficult to
experiment with directly. Examples include biofilm formation
\citep{johnson:2008}, radiative shock hydrodynamics \citep{goh2013prediction},
and the design of turbines \citep{Huang:2018}.  The approach
advocated by KOH models field-data from a physical
system a computer simulation plus an additional bias
correction (see our review in Section \ref{koh}). Computer
models are biased because they idealize physical dynamics and can have more
dials or knobs, so-called \emph{calibration parameters}, than
can be controlled in the field. 
Limitations on simulation budgets and field data necessitate meta-modeling.
KOH recommend coupled Gaussian
processes \citep[GPs,][]{williams2006gaussian} as (1) as surrogate
\citep{gramacy2020surrogates} for novel simulation, and (2) to learn an
appropriate bias correction.


Here we consider methods for efficient design of a simulation
experiment for the purpose of coping with a limited field experiment within
the context of KOH. Taken in isolation, design for GP
surrogates has a rich literature. Recipes range from
purely random to geometric space-fillingness, such as via Latin-Hypercube
sampling \citep[LHS,][]{mckay2000comparison} and minimax designs
\citep{johnson1990minimax}. Closed form analytics from GP posterior quantities
(again see Section \ref{koh}) may be leveraged to derive optimality criteria,
such as maximum entropy or minimum integrated mean-squared prediction
error (IMSPE), to develop designs \citep{sacks1989design}. These ideas may be
applied as one-shot, allocating runs all at once, or sequentially via active
learning \citep{seo2000gaussian}, which can offer an efficient approximation
due to submodularity \citep{wei2015submodularity} properties while hedging
against parametric specification of any (re-) estimated or fixed quantities.
This active/sequential approach is generally preferred when possible,
particularly for simulators with a lengthy computation time. Ultimately the
result is space-filling when variance/information criterion are measured
globally in the input space. For a more thorough review see, e.g.,
\citet[][Chapters 4--6]{gramacy2020surrogates}.

Literature on simulation design for improved field prediction 
within the coupled-GP KOH framework is more limited. Most are one-shot or are
focused on field design rather than computer model acquisition.
\citet{leatherman2017designing} built minimum IMSPE designs for combined field
and simulation data. \citet{arendt2016preposterior} used pre-posterior
analysis to improve identification via space-filling criterion.
\citet{krishna2021robust} proposed one-shot designs for physical
experimentation that is robust to modeling choices for bias correction.
\citet{williams2011batch} explored entropy and distance-based criteria in an
active learning setting for field experiments. \citet{morris2015physical}
similarly studied the selection of new field data sites, but in support of
computer model development.  \citet{castillo2019bayesian} use the
expected Shannon information gain to actively learn field data for the
purpose of calibration.  \citet{surer2023sequential} derive a criterion for
reducing functional uncertainty in the posterior of calibration parameters.
Here we treat unknown calibration parameters as a nuisance,
acknowledging that they comprise a primary interest in many
contexts \citep{bayarri2009modularization, Higdon:2004,
bryn2014learning,plumlee2017bayesian,gu2018jointly,tuo2015,tuo2016, wong2017,
plumlee2019,plumlee2017bayesian,gu2018scaled}, and instead focus on
accurate field prediction. In fact, none of these contributions address a
scenario where (new) field measurement is difficult, but new simulations can
be acquired to improve field predictions.

\citet{ranjan2011follow} provide some insight along those lines, comparing
reduction in field data IMSPE for surrogate-only designs. They found that new
batches of simulations should involve design inputs closely aligned with field
data, paired with random calibration input settings. They stopped short of
offering an automatable recipe for choosing new acquisitions for simulation
across both spaces simultaneously, possibly because they lacked
a closed form criterion that could easily be searched for new acquisitions.
More recently, \citet{chen2022apik} considered quadrature-based IMSPE in a
setup similar to KOH, but for coupled gradients. Lack of a closed form IMSPE
necessitated candidate-based, discrete optimization.  Although analytic
expressions for IMSPE have been developed in related contexts
\citep[e.g.,][]{leatherman2018computer, binois2019replication,
wycoff2021jcgs}, we are unaware of any targeting computer model runs for
improved KOH prediction. Having a closed form means that
closed-form derivatives can lend speed and stability to
searches for optimal acquisitions.

Here we prove a KOH-based IMSPE criterion, and
closed-form derivatives, and with those we reveal novel
insights about which additional simulations lead to improved prediction.
Rather than ``matching'' field data design inputs and being ``random'' on
calibration parameters \citep{ranjan2011follow}, we show that our criterion
balances new runs between calibration parameter
estimates and exploration of the remainder of the
input space. So KOH-IMSPE prefers to space-fill,
but highly values acquisitions near the current maximum
a-posteriori value of the calibration parameter. Similar, rule-of-thumb
analogues have been suggested recently. E.g., \citet{fer2018linking} acquired
batches of additional computer model data by mixing samples from the posterior
(90\%) of the calibration parameters with the prior (10\%).

The remainder of the paper is organized as follows. In Section \ref{revsec},
we review the elements in play: GPs, KOH, and sequential design. Our KOH-IMSPE
criterion is developed and explored in Section \ref{kohimspesec}. Section
\ref{implementation} provides implementation details and an empirical analysis
of KOH-IMSPE in a sequential design/active learning context. Section
\ref{reeapp} details a real-simulation/real data experiment
studying extraction of rare-Earth-elements (REEs) that motivated
this work. We conclude in Section \ref{discussec} with a brief discussion.

\hypertarget{revsec}{%
\section{Review of basic elements}\label{revsec}}

KOH calibration couples a GP surrogate with GP bias correction, and our
contribution involves active learning in this setting via IMSPE. These are
reviewed in turn with an eye toward their integration in Section~\ref{kohimspesec}.

\hypertarget{gpsec}{%
\subsection{Gaussian Process Regression}\label{gpsec}}

 A GP models
an \(N \times 1\) vector of univariate responses \(\bm{Y}_N = \bm{Y}(\bm{X}_N)\) at an
\(N\times d\) matrix of design of inputs \(\bm{X}_N\) as multivariate normal (MVN): \(\bm{Y}_N \sim \mathcal{N}_N (\bm{\mu}, \bm{\Sigma})\). In a regression
context, say as a surrogate for computer model
simulations \(\bm{Y}(\bm{X}_N)\), it is common to take \(\bm{\mu} = \bm{0}\) and move all of the
modeling ``action'' into the covariance structure \(\bm{\Sigma}\), which is
defined by the distances between rows of \(\bm{X}_N\). For example,\begin{equation}
\bm{Y}_N \sim \mathcal{N}_N\left(\bm{0}, \nu \bm{K}(\bm{X}_N)\right) \quad \mbox{where} \quad 
\bm{K}(\bm{X}_N)_{ij} = k(x_i, x_j) = \exp\left(-\sum_{l=1}^d\frac{(x_{il} - x_{jl})^2}{\theta_l}\right) + \delta_{(i=j)} g.
\label{eq:expkernel}
\end{equation} This specific choice of kernel \(k(\cdot, \cdot)\) and the
so-called ``hyperparameterization'' (via \(\nu\), \(g\) and \(\bm{\theta}\)) is meant
as an example only. Our contributions are
largely agnostic to these choices. When viewing \((\bm{Y}_N, \bm{X}_N)\) as training
data, the MVN in Eq. \eqref{eq:expkernel} defines a likelihood that can
be used to learn any unknowns \citep[see, e.g.,][]{williams2006gaussian, santner2018design, gramacy2020surrogates}. Often, computer model simulations are
deterministic, so the \emph{nugget} parameter \(g\) is
taken as zero (or small \(g  = \varepsilon > 0\) for better conditioned
\(\bm{K}(\bm{X}_N)\)).

Prediction for a new run \(\bm{x}\) is facilitated
by extending the MVN relationship in Eq. \eqref{eq:expkernel} to \(Y(\bm{x})\).
\begin{equation}
\begin{bmatrix}
Y(\bm{x}) \\
\bm{Y}_N
\end{bmatrix} \sim \mathcal{N}_{1 + N}\left(\begin{bmatrix}
0\\
\bm{0}
\end{bmatrix}, \nu \begin{bmatrix}
k(\bm{x}) & \bm{k}(\bm{x},\bm{X}_N)\\
\bm{k}(\bm{X}_N,\bm{x}) & \bm{K}(\bm{X}_N)
\end{bmatrix}\right) \label{eq:gpsim}
\end{equation}
Above, \(\bm{k}(\bm{x}, \bm{X}_N)\) provides cross-kernel
evaluations between rows of \(\bm{x}\) and \(\bm{X}_N\),
\(\bm{K}(\bm{X}_N) \equiv \bm{K}(\bm{X}_N,\bm{X}_N)\), and \(k(\bm{x}) \equiv k(\bm{x}, \bm{x})\). Then, standard MVN conditioning reveals
\(Y(\bm{x}) \mid \bm{Y}_N \sim \mathcal{N}_{1} (\hat{\mu}_N(\bm{x}), \hat{\Sigma}_N(\bm{x}))\), where
\begin{equation}
\hat{\mu}_N(\bm{x}) = \bm{k}(\bm{x},\bm{X}_N) \bm{K}(\bm{X}_N)^{-1}\bm{Y}_N  \quad\quad\quad
\hat{\Sigma}_N(\bm{x}) = \nu (k(\bm{x}) - \bm{k}(\bm{x},\bm{X}_N)\bm{K}(\bm{X}_N)^{-1}\bm{k}(\bm{X}_N,\bm{x})). \label{eq:krigvar}
\end{equation} These are known as the Kriging equations
\citep{matheron1963principles} in the geo-spatial literature, and
\(\hat{\mu}_N(\bm{x})\) can be shown to provide the best linear
unbiased predictor \citep{santner2018design}.

\begin{figure}[ht!]
\centering 
\includegraphics{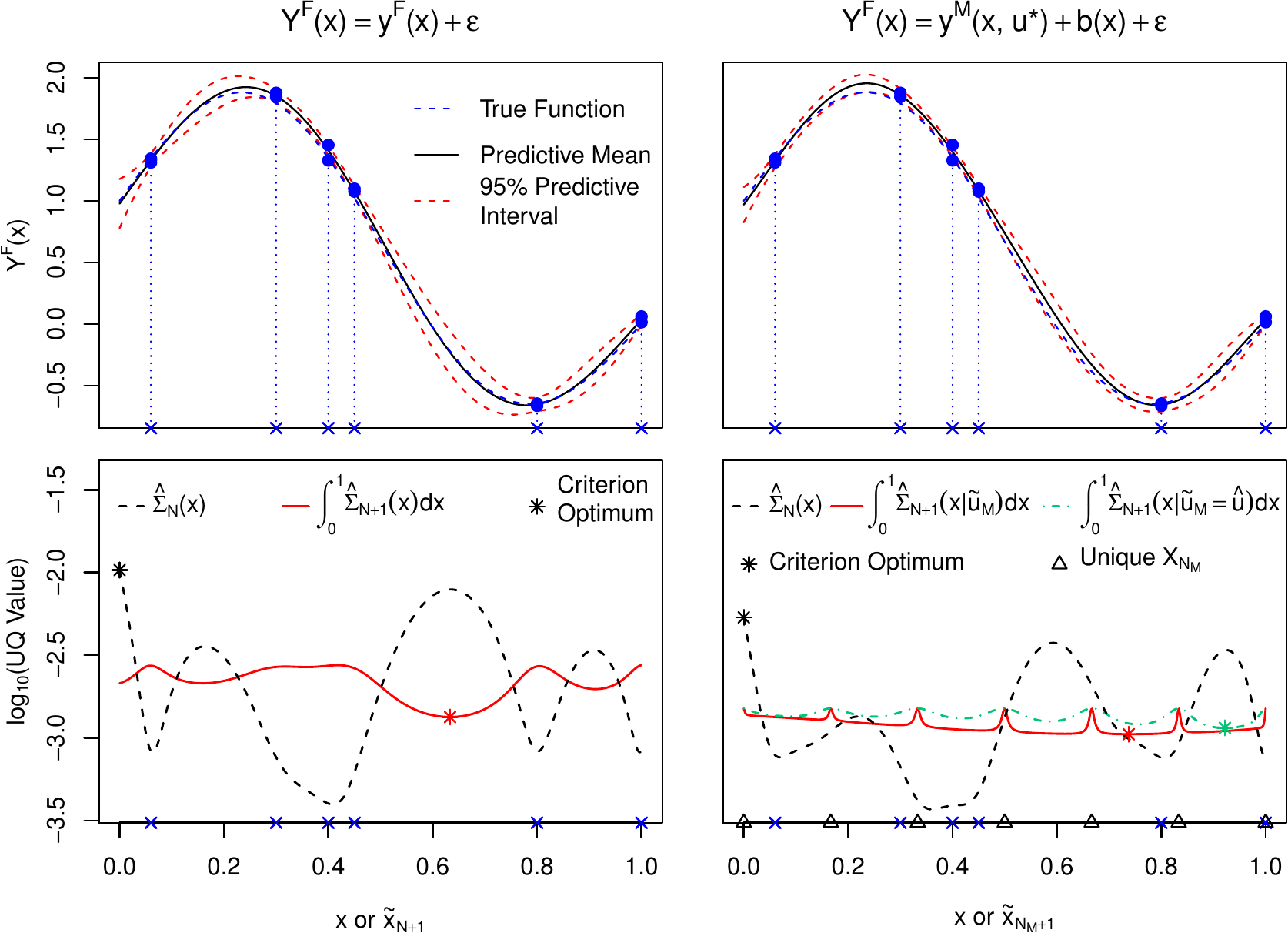} 
\caption{\textit{top panels:}  GP surfaces for field predictions
given just field data (\textit{left}) and combined with simulator
data using KOH (\textit{right}); \textit{bottom panels:} UQ and acquisition criterion corresponding to the panels
above.} \label{fig:predvarfig}
\end{figure}

The top left panel of Figure \ref{fig:predvarfig} shows the results of fitting
a zero mean GP to some example data, whose
details are delayed until
Section \ref{illustration}. The predictive mean and the true
mean function, solid black and dashed blue lines respectively, are
fairly close together, illustrating decent accuracy. Red dashed lines indicate
95\% predictive intervals, calculated over a fine grid from
\(\hat{\Sigma}_N(\bm{x})\) and shown as a dotted black line in the bottom
left panel. Observe that the interval becomes wider, depicting higher
uncertainty, in-between two training data locations.

\hypertarget{imspesec}{%
\subsection{Integrated Mean Squared Prediction Error}\label{imspesec}}

E.q.~\eqref{eq:krigvar} has many uses beyond
prediction, e.g., deducing where \(Y(\bm{x})\) is minimized via
Bayesian optimization \citep[BO,][]{jones1998efficient}; exploring which
coordinates of \(\bm{x}\) most influence \(Y(\bm{x})\)
\citep{marrel2009calculations}. We are interested in sequential experimental
design, or active learning, to select new runs for an improved fit/prediction. One idea is to choose \(\tilde{\bm{x}}_{N+1} =
\mathrm{argmax}_{\bm{x} \in \mathcal{X}} \;
\hat{\Sigma}_N(\bm{x})\), i.e., the location with maximum
predictive variance. Here \(\bm{x}\) and
\(\tilde{\bm{x}}_{N+1}\) represent a single \((N' = 1) \times d\) coordinate
vector in the input space \(\mathcal{X}\). \citet{mackay1992information}
showed that such acquisitions lead to (approximate) maximum entropy designs in
the context of neural networks, and \citet{seo2000gaussian} extended these to
GPs naming the idea \emph{active learning MacKay} (ALM). Despite their
simplicity, convenient closed form, and analytic gradients for efficient
library-based numerical optimization, ALM-based designs disappoint because they concentrate new runs on the boundary of
\(\mathcal{X}\). 

As a remedy, \citet{sacks1989design} proposed an integrated
mean-squared prediction error (IMSPE) criterion: \(\mathrm{IMSPE}(\bm{X}_N) =
\int_{\mathcal{X}} \hat{\Sigma}_N(\bm{x}) \; d\bm{x}\). Again
\(\bm{x}\) is \(d \times 1\) so that the integral is
\(d\)-dimensional. After conditioning on
hyperparameters, IMSPE is a deterministic, scalar function of \(\bm{X}_N\).
One could use IMSPE to determine an entire
\(N \times d\) design \(\bm{X}_N\) in one shot by optimizing over all \(Nd\)
coordinates as in
\(\tilde{\bm{X}}_N = \mathrm{argmin}_{\bm{X}_N \in \mathcal{X}^N} \mathrm{IMSPE}(\bm{X}_N)\), or
to augment \(\bm{X}_N\) with a new row \(\tilde{\bm{x}}_{N+1} = \mathrm{argmin}_{\bm{x}_{N + 1} \in \mathcal{X}}\mathrm{IMSPE}([\bm{X}_N; \bm{x}_{N+1}^{\top}])\) in
an active learning setting. Independently, \citet{cohn1994neural} developed a
similar criterion for neural networks,
 approximating the integral as a sum; \citet{seo2000gaussian}
extended this to GPs, coining Active Learning Cohn (ALC).

Here we follow the mathematics laid out by \citet{binois2019replication}, who
provided a closed form IMSPE and gradient under a uniform measure for
\(\mathcal{X} \in [0,1]^d\), i.e.~\(p(\mathcal{X}) = \mathcal{U} [0,1]^d\). 
The approach is at once elegant and practical in implementation,
and consequently has spurred a cottage industry of variations
\citep{wycoff2021jcgs, cole2021locally, sauer2021active} of which our main contribution can be viewed as yet another.
The derivation relies on two trace identities: \(\mathrm{tr}(\bm{ABC}) = \mathrm{tr}(\bm{BCA})\); and \(\mathrm{tr}(\bm{A}_{m\times m}\bm{B}_{m\times m}) = \bm{1}^\top (\bm{A}_{m\times m} \circ \bm{B}_{m\times m})\bm{1}\) where \(\bm{1}\) is a vector of ones with
a length equal to \(m\), and \(\circ\) is the Hadamard, or element wise, product. It also involves a re-positioning of the integral inside of the
matrix trace, which is legitimate as both are linear operators. A uniform
measure yields closed forms for \(\bm{W}(\bm{X}_N) = \bm{W}(\bm{X}_N,\bm{X}_N) = \int_{[0,1]^d} \bm{k}(\bm{X}_N,\bm{x}) \bm{k}(\bm{x},\bm{X}_N)^\top \; d\bm{x}\) as an \(N \times N\) under common kernels \(k(\cdot, \cdot)\),
which are not duplicated here. See the appendix of \citet{binois2019replication}. Combining those elements, and interpreting the integral as an expectation
over uniform \(\bm{x}\), yields\begin{equation}
\begin{split}
\mathrm{IMSPE}(\bm{X}_N) &= \int_{[0,1]^d} \nu k(\bm{x}) - \nu \bm{k}(\bm{x},\bm{X}_N) \bm{K}(\bm{X}_N)^{-1}\bm{k}(\bm{X}_N,\bm{x}) \; d\bm{x}\\
&= \nu - \int_{[0,1]^d} \mathrm{tr}\left( \nu \bm{K}(\bm{X}_N)^{-1}  \bm{k}(\bm{X}_N,\bm{x}) \bm{k}(\bm{x},\bm{X}_N) \right) \; d\bm{x}\\
&= \nu - \nu \bm{1}^\top \left( \bm{K}(\bm{X}_N)^{-1} \circ \bm{W}(\bm{X}_N)  \right) \bm{1}.
\end{split}
\label{eq:imspe}
\end{equation}We introduce these techniques, and enhanced level of detail for a review,
because our own work in Section \ref{kohimspesec} involves similar operations.
In practice, estimates of \(\nu\) tacitly condition on \((\bm{X}_N,\bm{Y}_N)\), however the typical development presumes
that we don't have \(\bm{Y}\)-values yet, or at least not \(y_{N+1}\) in the
active learning context. Consequently, it is equivalent to choose
\begin{equation}
\tilde{\bm{x}}_{N+1} = \underset{\bm{x}_{N + 1} \in [0,1]^d}{\mathrm{argmax}} \  \bm{1}^\top \left( \bm{K}(\bm{X}_{N+1})^{-1} \circ \bm{W}(\bm{X}_{N+1})  \right) \bm{1} \quad\quad \mbox{ where } \bm{X}_{N+1} \equiv [\bm{X}_N; \bm{x}_{N+1}^\top],
\label{eq:minimspe}
\end{equation} 
i.e., where \(\bm{X}_N\) is augmented with the new row \(\bm{x}_{N+1}^\top\).
In this way, the acquisition explicitly targets predictive accuracy by
minimizing mean-squared prediction error. 

Returning to Figure \ref{fig:predvarfig}, the bottom left plot allows for comparison of the dotted black line plotting predictive variance, and the solid red line plotting IMSPE as a function of the \(\bm{x}\) augmenting the initial design. Observe that the minimum IMSPE point \(\tilde{\bm{x}}_{N+1}\), a red asterisk, does not maximize the predictive variance.

\hypertarget{koh}{%
\subsection{Simulator calibration}\label{koh}}

KOH propose that field observations
\(\bm{Y}^F(\bm{X}_{N_F})\), at $p$-dimensional inputs \(\bm{X}_{N_F}\),
be modeled as noisy realizations of a computer model \(y^M(\bm{X}_{N_F},
\bm{U}_{N_F \times s})\) set at the ideal/true value of \(\bm{U}_{N_F} =
\bm{U}^{\star} = \bm{1}_{N_F}(\bm{u}^\star)^{\top}\), modulo an additive bias correction
\(b(\bm{X}_{N_F})\), representing any systematic discrepancy inherent in the
computer model: $
\bm{Y}^F(\bm{X}_{N_F}) = \bm{y}^M(\bm{X}_{N_F},\bm{U}^\star) + \bm{b}(\bm{X}_{N_F}) + \bm{\epsilon}$ where $\bm{\epsilon} \sim \mathcal{N}_{N_F}(0, \pmb{\mathbb{I}}_{N_F} \sigma^2)$.
 We follow \citet{bayarri2009modularization}'s ``modularized''
update to KOH and place an independent GP prior on \(\bm{b}(\cdot)\) so that
\(\hat{\bm{b}}(\bm{X}_{N_F})\) may be estimated based on observed
discrepancies \(\bm{b}(\bm{X}_{N_F}) = \bm{y}^F - \bm{y}^M(\bm{X}_{N_F},
\hat{\bm{U}})\).  Estimates for \(\hat{\bm{u}}\) and noise \(\hat{\sigma}^2\)
may be obtained jointly with \(\hat{\bm{b}}(\cdot)\). Under an
independent GP prior for \(\bm{Y}^M(\bm{x},\bm{u})\), the joint marginal
likelihood, governing prediction and hyperameter inference, for
field data
\(\bm{Y}_{N_F}\), computer model runs \(\bm{Y}_{N_M}\), and predictive outputs \(Y^F(\bm{x}) \equiv Y^F(\bm{x}, \hat{\bm{u}})\) derives from an MVN of dimension
\(1+N_F + N_M\) with mean zero and covariance matrix
\begin{equation}
\mathbb{C}\mathrm{ov}\left(
\begin{bmatrix}
Y^F(\bm{x}, \hat{\bm{u}})\\
\bm{Y}_{N_F}\\
\bm{Y}_{N_M}
\end{bmatrix}
\right) = \bm{\Sigma}^{M}+ \bm{\Sigma}^B
\quad \mbox{where} \quad 
\bm{\Sigma}^B
 = \nu_B \begin{bmatrix}
k^B{(\bm{x})} &   \bm{k}^B{(\bm{x},\bm{X}_{N_F})} & \bm{0}^{\top}_{N_M} \\
\bm{k}^B{(\bm{X}_{N_F}, \bm{x})} & \bm{K}^B(\bm{X}_{N_F}) & \bm{0}_{N_F \times N_M} \\
\bm{0}_{N_M} & \bm{0}_{N_M \times N_F} & \bm{0}_{N_M \times N_M}\\
\end{bmatrix}
\label{eq:kohimspecov}
\end{equation}
\begin{equation}
\mbox{and} \quad \bm{\Sigma}^M = \nu_M\begin{bmatrix}
k([\bm{x}, \hat{\bm{u}}]) &   \bm{k}{([\bm{x}, \hat{\bm{u}}],[\bm{X}_{N_F}, \hat{\bm{U}}])} &  \bm{k}{([\bm{x}, \hat{\bm{u}}],[\bm{X}_{N_M}, \bm{U}_{N_M}])}\\
\bm{k}{([\bm{X}_{N_F}, \hat{\bm{U}}], [\bm{x}, \hat{\bm{u}}])} & \bm{K}([\bm{X}_{N_F}, \hat{\bm{U}}]) & \bm{K}([\bm{X}_{N_F}, \hat{\bm{U}}],[\bm{X}_{N_M}, \bm{U}_{N_M}])\\
\bm{k}([\bm{X}_{N_M}, \bm{U}_{N_M}],[\bm{x}, \hat{\bm{u}}]) & \bm{K}([\bm{X}_{N_M}, \bm{U}_{N_M}],[\bm{X}_{N_F}, \hat{\bm{U}}]) & \bm{K}([\bm{X}_{N_M}, \bm{U}_{N_M}])
\end{bmatrix}.
\notag
\end{equation}
Above, subscripts \(\nu_M\) and \(\nu_B\) indicate separate scale
hyperparameters for surrogate and bias GPs, respectively. We use
\(\bm{k}(\cdot)\) and \(\bm{K}(\cdot)\) for the computer model kernel and
 introduce \(B\) superscripts for the bias analog. For \(\bm{Y}_{N_F}\), the
 calibration parameter estimate is included as \(\bm{U} = \bm{1}_{N_F}
 \hat{\bm{u}}^{\top}\). Implicit in
\(\bm{K}^B(\cdot)\) is an additive diagonal \(\pmb{\mathbb{I}_{N_F}}g\) so
that a nugget parameter \(g\) can capture noise with variance \(\sigma^2 =
\nu_B g\).  
%
 Predictions follow Eq.~\eqref{eq:krigvar}, except 
except with more elaborate conditioning. The mean is not central to our
discussion here,
$\hat{\Sigma}(\bm{x} \mid \hat{\bm{u}}) \equiv 
\mathbb{V}\mathrm{ar} ( Y^F(\bm{x},\hat{\bm{u}}) \mid \bm{Y}_{N_F}, \bm{Y}_{N_M})$ is
important for calculating
IMSPE. 
\begin{align}
\hat{\Sigma}(\bm{x} \mid \hat{\bm{u}}) 
\label{eq:krigvarfield} 
&= \nu_M k([\bm{x}, \hat{\bm{u}}])  + \nu_B k^B{(\bm{x})} - \begin{bmatrix}
\nu_M \bm{k} + \nu_B \bm{k}^B
\end{bmatrix}^\top \begin{bmatrix}
 \bm{\Sigma}^{M,B}_{N_F + N_M}
\end{bmatrix}^{-1}
\begin{bmatrix}
\nu_M \bm{k} + \nu_B \bm{k}^B
\end{bmatrix} \\
\mbox{where} \quad \bm{k} &= \begin{bmatrix}
\bm{k}{([\bm{X}_{N_F}, \hat{\bm{U}}], [\bm{x}, \hat{\bm{u}}])} \\
\bm{k}([\bm{X}_{N_M}, \bm{U}_{N_M}],[\bm{x}, \hat{\bm{u}}])
\end{bmatrix} \quad\quad\quad 
\bm{k}^B = \begin{bmatrix}
\bm{k}^B{(\bm{X}_{N_F}, \bm{x})} \\
\bm{0}_{N_M}
\end{bmatrix} \notag \\
\bm{\Sigma}^{M,B}_{N_F + N_M} &=  \begin{bmatrix}
\nu_M \bm{K}([\bm{X}_{N_F}, \hat{\bm{U}}]) + \nu_B \bm{K}^B(\bm{X}_{N_F}) & \nu_M \bm{K}([\bm{X}_{N_F}, \hat{\bm{U}}],[\bm{X}_{N_M}, \bm{U}_{N_M}])\\
\nu_M \bm{K}([\bm{X}_{N_M}, \bm{U}_{N_M}],[\bm{X}_{N_F}, \hat{\bm{U}}]) & \nu_M \bm{K}([\bm{X}_{N_M}, \bm{U}_{N_M}])
\end{bmatrix}. \notag
\end{align}
The top right panel of Figure \ref{fig:predvarfig} shows a GP fit using the
same data in the top left plot, but augmented with 49 simulator runs
(\(\bm{Y}_{N_M}, \bm{X}_{N_M}, \bm{U}_{N_M}\)), on
an evenly spaced grid in \([0,1]^2\), using the KOH framework. This extra
information results in reduced predictive variance across the
input space as indicated by comparing black dotted lines in the bottom panels.
More data, even if the data is not field data, reduces predictive uncertainty.
Further reductions could be realized with even more computer model runs, which
is the target of our main methodological contribution.

\hypertarget{kohimspesec}{%
\section{Optimal acquisition of new simulator runs}\label{kohimspesec}}

Here we derive a closed form active learning criterion by deploying IMSPE in
the KOH framework. Gradients are provided to facilitate efficient
and stable numerical
optimization. These also yield additional insight into the value of potential
new simulation runs.

\hypertarget{kohderiv}{%
\subsection{Closed form KOH-IMSPE and derivative}\label{kohderiv}}

Predictive variance \(\hat{\Sigma}(\bm{x} \mid \hat{\bm{u}})\)
\eqref{eq:krigvarfield} in hand, the key step in deriving a
KOH-IMSPE acquisition criterion is to integrate over the input space of the
field predictive location \(\bm{x}\). Throughout this discussion we condition
the evaluation of KOH-IMSPE and its gradient on point estimates
of \((\hat{\bm{u}}, \hat{\nu}_M, \hat{\nu}_B), \bm{\theta}^M, \bm{\theta}^B\)
and \(g\). The maximum a-posteriori \citep[MAP; e.g.,][Section
9.1]{gramacy2020surrogates} of each hyperparameter is evaluated and updated
after each active learning acquisition, separate from the
KOH-IMSPE integral \eqref{eq:kohimspefinal}. Our focus here is 
on how to acquire new simulator runs given those values. Section
\ref{discussec} offers thoughts on how alternative hyperparameter estimation
strategies might impact our active learning setup.

The series of equations below outline our approach to integrating 
\(\hat{\Sigma}(\bm{x} \mid \hat{\bm{u}})\) over \(\bm{x}\), beginning with Eq. \eqref{eq:krigvarfield}. Hats are dropped on \(\nu_M\) and \(\nu_B\) to reduce clutter. Although expressed here as a function of the computer model design \([\bm{X}_{N_M + 1}, \bm{U}_{N_M + 1}]\), IMSPE would also depend upon the field data.
\begin{align}
\mathrm{IMSPE}&([\bm{X}_{N_M + 1}, \bm{U}_{N_M + 1}]) = \int_{[0,1]^p} \hat{\Sigma}(\bm{x} \mid \hat{\bm{u}}) \; d\bm{x} \notag \\
&= \nu_M + \nu_B - \int_{[0,1]^p} \mathrm{tr}\left( \left[\bm{\Sigma}^{M,B}_{N_F + N_M + 1}\right]^{-1} \right. \notag \\
& \qquad \left. \vphantom{\left[\bm{\Sigma}^{M,B}_{N_F + N_M + 1}\right]^{-1}} \left(\nu_M^2\bm{k}\bm{k}^\top + 2\nu_M \nu_B \bm{k}^B \bm{k}^\top  + \nu_B^2 \bm{k}^B [\bm{k}^B]^\top\right)\right) \; d\bm{x} \label{eq:kohimspemid}\\
&= \nu_M + \nu_B - \int_{[0,1]^p} \bm{1}^\top \left(\left[\bm{\Sigma}^{M,B}_{N_F + N_M + 1}\right]^{-1}\circ \right. \notag \\
& \qquad \left. \vphantom{\left[\bm{\Sigma}^{M,B}_{N_F + N_M + 1}\right]^{-1}} \left(\nu_M^2\bm{k}\bm{k}^\top + 2\nu_M \nu_B \bm{k}^B \bm{k}^\top  + \nu_B^2 \bm{k}^B [\bm{k}^B]^\top\right)\right) \bm{1} \; d\bm{x} \label{eq:kohimspehadamard}\\
&= \nu_M + \nu_B - \bm{1}^\top \left(\left[\bm{\Sigma}^{M,B}_{N_F + N_M + 1}\right]^{-1} \circ\right. \notag \\
& \qquad \left. \vphantom{\left[\bm{\Sigma}^{M,B}_{N_F + N_M + 1}\right]^{-1}} \left(\nu_M^2 \bm{W}^{M,M} +  2 \nu_M \nu_B \bm{W}^{M,B} +  \nu_B^2 \bm{W}^{B,B}\right)\right)\bm{1}
\label{eq:kohimspefinal}
\end{align}
Trace identities \(\mathrm{tr}(\bm{ABC}) = \mathrm{tr}(\bm{BCA})\) and \(\mathrm{tr}(\bm{A}^{\top}) = \mathrm{tr}(\bm{A})\) are involved in \eqref{eq:kohimspemid}, whereas \eqref{eq:kohimspehadamard} utilizes \(\mathrm{tr}(\bm{AB}) = \bm{1}^\top (\bm{A} \circ \bm{B})\bm{1}\). Notice that \(\hat{\Sigma}(\bm{x} \mid \hat{\bm{u}})\) has been broken into a weighted sum of functions specifying the covariance between \(\bm{x}\) and the observations, providing a less abstract format for integration. Integration produces \eqref{eq:kohimspefinal}, where for any \(\{\alpha, \beta\} \in \{M,B\}\): \(\bm{W}^{\alpha, \beta} = \int_{[0,1]^p} \bm{k}^\alpha (\bm{k}^\beta)^\top \; d\bm{x}\). Solutions for the elements of \(\bm{W}^{\cdot, \cdot}\) are kernel-dependent. Supplementary Material \ref{intA} provides forms compatible with Gaussian kernels. Similar derivations for Matérn kernels can be found in \citet{binois2019replication}.

Acquiring a new computer model run \([\tilde{\bm{x}}, \tilde{\bm{u}}]\) requires minimizing IMSPE \eqref{eq:kohimspefinal} with the augmented design:
\begin{equation}
[\tilde{\bm{x}},\tilde{\bm{u}}] = \! \underset{[\bm{x},\bm{u}]_{N_M + 1} \in \mathcal{X}^d}{\mathrm{argmin}} \! \mathrm{IMSPE}([\bm{X}_{N_M + 1}, \bm{U}_{N_M +1}]), \; \mbox{where} \; [\bm{X}_{N_M + 1}, \bm{U}_{N_M + 1}] \equiv [\bm{X}_{N_M}, \bm{U}_{N_M};[\bm{x},\bm{u}]_{N_M + 1}].
\label{eq:prog}
\end{equation}
To continue with our running illustration, the bottom-right panel of Figure \ref{fig:predvarfig} shows KOH-IMSPE instead of typical IMSPE. Notably, IMSPE, along with predictive variance, is lower for the KOH GP fit. Because of the additional computer model data, the KOH-IMSPE surface is still multi-modal, but flat relative to the GP fit with only field data (bottom-left panel). Additionally, the point which minimizes IMSPE when computer simulator data is introduced is different than when no computer simulator data is used.

Considering the multi-modality of the KOH-IMSPE acquisition surface, and its
relative flatness compared with ordinary IMSPE, a thoughtful strategy for
solving the program in Eq. \eqref{eq:prog} is essential to obtaining good
acquisitions. Local numerical optimization via finite differentiating in this
setting can be fraught with numerical challenges. One option is to deploy a
discrete candidate set for \(\tilde{\bm{X}_N}\), such as via LHS or
triangulation \citep{gramacy2022triangulation}. Our analytical solution of
KOH-IMSPE \eqref{eq:kohimspefinal} allows closed form
gradient-based minimization (e.g., BFGS, \citet{byrd1995limited}).

The derivative of KOH-IMSPE \eqref{eq:kohimspefinal} with
respect to element \(l\) of \([\tilde{\bm{x}}, \tilde{\bm{u}}]\) is shown in
Eq. \eqref{eq:imspediff}. Since \([\tilde{\bm{x}}, \tilde{\bm{u}}]\) is
contained within \(\bm{\Sigma}^{M,B}, \bm{W}^{M,M},\) and \(\bm{W}^{M,B}\), we
need a chain rule for matrices and inverses:
\[
\frac{\partial(\bm{U}(x)\circ \bm{V}(x))}{\partial x} = 
\bm{U}(x)\circ\frac{\partial \bm{V}(x)}{\partial x} + \frac{\partial \bm{U}(x)}{\partial x}\circ 
\bm{V}(x) \quad \mbox{ and } \frac{\partial \bm{U}(x)^{-1}}{\partial x} =
 -\bm{U}(x)^{-1}\frac{\partial  \bm{U}(x)}{\partial x}\bm{U}(x)^{-1}.
\]
Using these  identities, we find that the derivative of KOH-IMSPE can be written as follows:
\begin{align}
\frac{\partial \, \mathrm{IMSPE}}{\partial  [\tilde{\bm{x}}, \tilde{\bm{u}}]_l} &= 
\bm{1}^{\top}\!\left(\!\left[\bm{\Sigma}^{M, B}_{N_F + N_M +1}\right]^{-1}
\frac{\partial \bm{\Sigma}^{M, B}_{N_F + N_M + 1}}{\partial [\tilde{\bm{x}}, 
\tilde{\bm{u}}]_l}\left[\bm{\Sigma}^{M, B}_{N_F + N_M + 1}\right]^{-1}\!\!\!\!\!\circ 
\left(\nu_M^2 \bm{W}^{M,M} \!+\! 2\nu_M\nu_B \bm{W}^{M,B} \!+\! \nu_B^2 \bm{W}^{B,B}  \right)\right. - \notag \\
& \qquad \left. \left[\bm{\Sigma}^{M, B}_{N_F + N_M + 1}\right]^{-1}\!\!\!\circ \left( \nu_M^2 
\frac{\partial \bm{W}^{M,M}}{\partial[\tilde{\bm{x}}, \tilde{\bm{u}}]_l} + 2\nu_M\nu_B 
\frac{\partial \bm{W}^{M,B}}{\partial[\tilde{\bm{x}}, \tilde{\bm{u}}]_l} \right)\!\right)\bm{1},
\label{eq:imspediff}
\end{align}
where forms for \(\frac{\partial \bm{W}^{M,M}}{\partial
[\tilde{\bm{x}},\tilde{\bm{u}}]}\), \(\frac{\partial \bm{W}^{M,B}}{\partial
[\tilde{\bm{x}},\tilde{\bm{u}}]}\), and \(\frac{\partial\bm{\Sigma}^{M,
B}_{N_F + N_M + 1} }{\partial[\tilde{\bm{x}}, \tilde{\bm{u}}]}\) are provided
in Supplements \ref{intA} and \ref{blockmatrixA}.

It is illuminating to study how this gradient behaves when \(\tilde{\bm{u}} =
\hat{\bm{u}}\).  The second term in the sum in
Eq.~\eqref{eq:imspediff}, related to \(\frac{\partial \bm{W}^{M,M}}{\partial
[\tilde{\bm{x}},\tilde{\bm{u}}]_l}\) and \(\frac{\partial
\bm{W}^{M,B}}{\partial [\tilde{\bm{x}},\tilde{\bm{u}}]_l}\), is a matrix of
zeros when differentiating with respect to \(\tilde{\bm{u}}\) at
\(\tilde{\bm{u}} = \hat{\bm{u}}\).  The location of the zero-valued gradient
for KOH-IMSPE in \(\bm{U}\)-space represents a balance point in the
correlation between how close \([\bm{x}, \hat{\bm{u}}]\) is to
\([\tilde{\bm{x}}, \tilde{\bm{u}}]\), and how close \([\bm{X}_{N_M},
\bm{U}_{N_M}]\) is to \([\tilde{\bm{x}}, \tilde{\bm{u}}]\).
Diving into the first term, the differential covariance matrix
with respect to \(\tilde{\bm{u}}\) at \(\tilde{\bm{u}} = \hat{\bm{u}}\), when
the Gaussian kernel is used, is
\begin{equation}
\begin{split}
&\frac{\partial \bm{\Sigma}^{M,B}}{\partial \tilde{u}_l}:(\tilde{u}_l = \hat{u}_l) = \nu_M \times \\
& \begin{bmatrix}
\bm{0}_{N_F \times N_F} & \bm{0}_{N_F \times N_M} & \bm{0}_{N_F}\\
\bm{0}_{N_M \times N_F} & \bm{0}_{N_M \times N_M} & 
\left(\frac{\partial \bm{k}([\bm{X}_{N_M},\bm{U}_M], 
[\tilde{\bm{x}}, \tilde{\bm{u}}]) }{\partial \tilde{u}_l}\right)_{N_M}\\
\bm{0}^{\top}_{N_F} & 
\left(\frac{\partial \bm{k}([\bm{X}_{N_M},\bm{U}_M], 
[\tilde{\bm{x}}, \tilde{\bm{u}}]) }{\partial \tilde{u}_l}\right)^{\top}_{N_M} & 0_{1\times 1}
\end{bmatrix}.
\end{split}
\label{eq:dkdutilde}
\end{equation}
Notably the matrix in Eq.~\eqref{eq:dkdutilde} is mostly, but not
completely, zero since \(\partial \bm{k}([\tilde{\bm{x}},
\tilde{\bm{u}}],[\bm{X}_{N_F}, \hat{\bm{U}}]\mid \tilde{\bm{u}} =
\hat{\bm{u}})/\partial \tilde{u}_l \propto
-\frac{2(\tilde{u_l}-\hat{u_l})}{\theta_l^M}\exp\left(-(\hat{u}_l-\tilde{u}_l)^2/\theta_l^M\right)
= \bm{0}\).  Within the gradient, this sparse differential
covariance matrix is weighted by the squared inverse covariance matrix, and
the correlation between the design and the integrated predictive location.
For this special case, the KOH-IMSPE gradient is only zero when
\[
\bm{1}^{\top}\left(\frac{\partial \bm{\Sigma}^{M, B}_{N_F + N_M +
1}}{\partial [\tilde{\bm{x}}, \tilde{\bm{u}}]_{l }}\circ\left[\bm{\Sigma}^{M,
B}_{N_F + N_M + 1}\right]^{-1} \left(\nu_M^2 \bm{W}^{M,M} + 2\nu_M\nu_B
\bm{W}^{M,B} + \nu_B^2 \bm{W}^{B,B}  \right)\left[\bm{\Sigma}^{M, B}_{N_F +
N_M +1}\right]^{-1}\right)\bm{1} 
\]
evaluates to zero.
Any trace of the multiplication of the differential covariance matrix by
another matrix will be near, but  importantly not
guaranteed to be, at minimum when
\(\tilde{\bm{u}}=\hat{\bm{u}}\). This result reveals a trade-off between
exploratory behavior, away from \(\hat{\bm{u}}\), and exploitation nearby
\(\hat{\bm{u}}\). 

\hypertarget{illustration}{%
\subsection{Illustration}\label{illustration}}

\begin{figure}[ht!]
\includegraphics{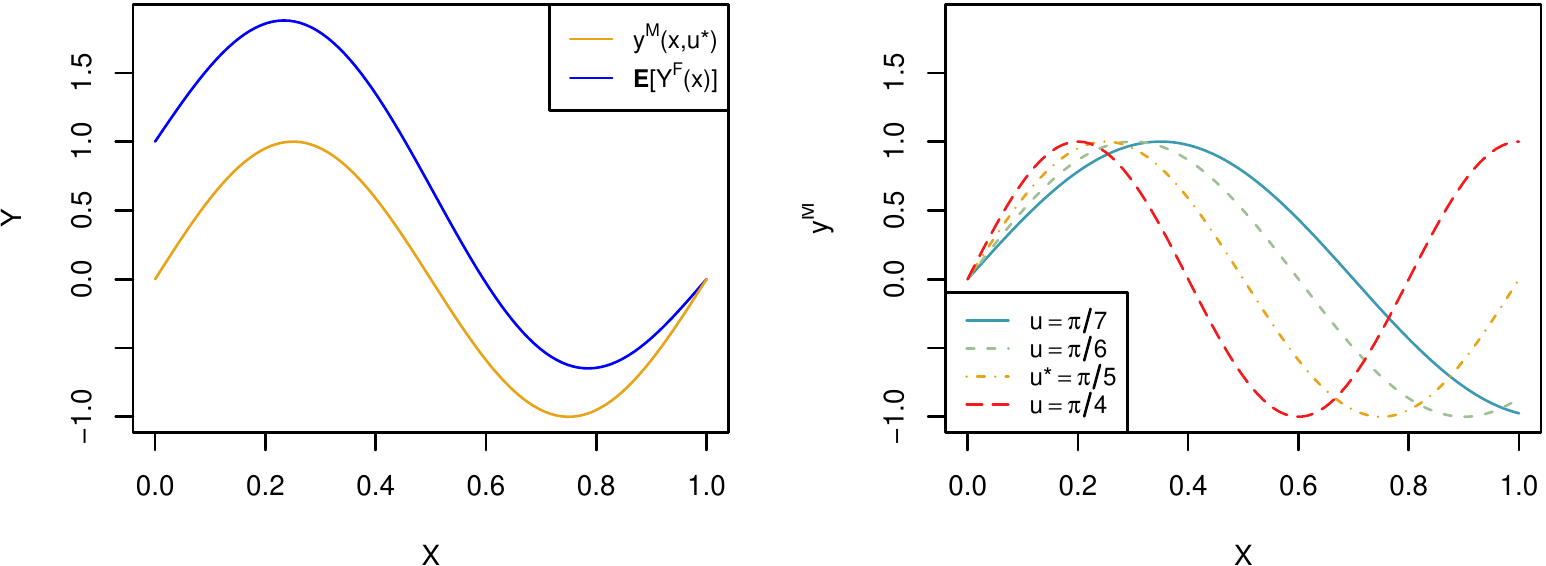} \caption{True surface, model surface with $u^\star$, and effect of varying $u$ on the model surface.}\label{fig:sinfns}
\end{figure}

Consider the following data-generating mechanism with 1d design input \((x)\)
and 1d calibration parameter \((u)\).
\begin{align*}
y^M(x,u) &= \sin (10x, u) &
b(x) &= 1-\frac{1}{3}x - \frac{2}{3}x^2 \\
Y^F(x) &= y^M\left(x, u^\star = \frac{\pi}{5}\right) + b(x) + \epsilon 
& \epsilon &\stackrel{\mathrm{iid}}{\sim} \mathcal N(0, 0.1^2)
\end{align*}
This setup underpins the running illustrative example supported by
Figure \ref{fig:predvarfig}. The left panel of
Figure \ref{fig:sinfns} shows the mean of the field data \(\mathbb{E}(Y^F(x)) = 
y^M\!\left(x, u^\star = \frac{\pi}{5}\right) + b(x)\) juxtaposed against
the computer model \(y^M\!\left(x, u^\star = \frac{\pi}{5}\right)\), whereas
the right panel shows how other settings of \(u\) affect that computer model.

Here we explore KOH-IMSPE on this example. To initialize, a \(N_M = 10\)-sized
2d random LHS was used for the initial computer model design in
\(\lbrack \bm{X}_N, \bm{U}\rbrack\)-space. Field data were collected as two replicates
of five unique locations (i.e., \(N_F = 10\)) on an equally spaced grid on
\(\mathcal{X}\). Following \citet{bayarri2009modularization} and
\citet[][Section 9.2]{gramacy2020surrogates}, we fit a GP
surrogate to the simulator data and bias GP to the residuals from the field
data, thereby estimating \(\hat{u}\) jointly with other hyperparameters via
MAP with \(p(u) = \mathrm{Beta}(2,2)\), a standard regularizing prior on the
calibration parameter. We then proceeded with 25 KOH-IMSPE acquisitions of new
computer model runs. After each acquisition the model(s) were re-fit to
prepare for the next one.

\begin{figure}[ht!]

{\centering \includegraphics{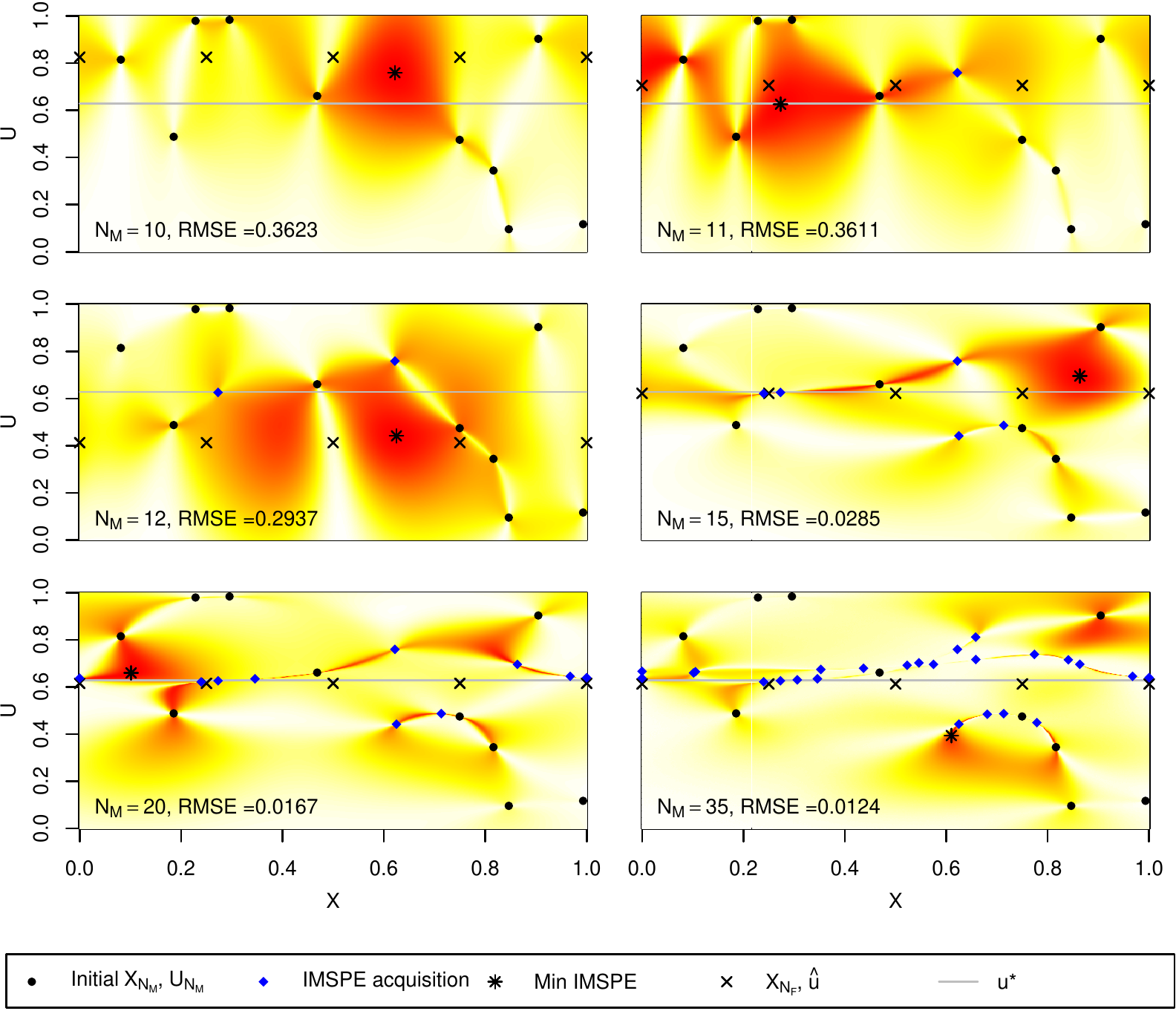} 

}

\caption{KOH-IMSPE surface in $X, U$ space as points are sequentially added to an initial computer model design.  Red indicates lower values and white/yellow indicates larger values.}\label{fig:twodfig}
\end{figure}

Figure \ref{fig:twodfig} shows KOH-IMSPE surface plots for \(N_M =\)
10, 11, 12, 15, 20, 35, the location of minimum KOH-IMSPE, the initial
computer model design, and the points previously added via KOH-IMSPE. Red colors in the surface indicate smaller KOH-IMSPE values. The location of
\(u^\star\) is shown as a grey line, which is fixed in each panel. Field data locations are plotted at \((\bm{X}_{N_F}, \hat{u})\) for the estimate of \(\hat{u}\) found at the given size of \(N_M\), which is distinct in each panel.  Root mean squared error (Eq. \eqref{eq:rmse}) on a noiseless testing set is shown at each step.
Observe that minimum KOH-IMSPE, i.e., \(\tilde{u}\) solving Eq. \eqref{eq:prog} is often found near \(\hat{u}\), but not always. Distinct exploratory behavior is still observed in the \(u\)-coordinate, i.e., reflecting the tradeoff between exploration and exploitation suggested by the analysis of derivative information at the end of Section \ref{kohderiv}. There are also a diversity of \(x\)-values, marginally.

\hypertarget{implementation}{%
\section{Implementation and benchmarking}\label{implementation}}

Code for all illustrative and
benchmarking examples herein is in \textsf{R} and may be found in our Git
repository \href{https://bitbucket.org/gramacylab/kohdesign}{\texttt{https://bitbucket.org/gramacylab/kohdesign}}. Subroutines found therein liberally borrow
from \texttt{laGP} \citep{gramacy2016lagp} and \texttt{hetGP} \citep{hetGP}
libraries to make predictions, find MAP estimates of kernel hyperparameters (under Gamma priors detailed later), to build
covariance matrices, and evaluate integrals when appropriate. Throughout, inputs \([\bm{X}_N,\bm{U}]\) are scaled to \([0,1]\) to improve numerical stability and simplify code for the required integrals. Independent \(\mathrm{Beta}(2,2)\) priors were  placed on each element of \(\mathbf{u}\), and MAP solutions under a modularized KOH \citep{bayarri2007framework} yielded \(\hat{\mathbf{u}}\)  via
\cite{nelder1965simplex} or \citet{brent2013algorithms} as described by \citet{gra:etal:2015}
and demonstrated in \citet{gramacy2020surrogates}, Section 9.1.

Our implementation of the numerical calculation of the closed form of Eq.~\eqref{eq:kohimspefinal} is faithful to the description detailed in Supplement \ref{intA}.   However, without careful implementation, a sequential KOH-IMSPE design can be slow and unstable.
To enhance stability, and reduce computation time involved in the search
of the next acquisition, \([\tilde{\bm{x}}, \tilde{\bm{u}}]\) solving
\eqref{eq:minimspe} utilized closed form gradients (\ref{eq:imspediff}). These are also furnished
in \ref{intA}, including gradients of all \(\bm{W}^{\cdot, \cdot}\) values,
as required. To further reduce computation time, block matrix
inversion \citep{bernstein2009matrix} of \(\bm{\Sigma}^{M,B}\) was utilized to
avoid full covariance matrix inversion for every candidate entertained
by the optimizer. However, block matrix inversion substantially complicates the expressions for gradients. These
additional details are provided in
Supplement \ref{blockmatrixA}. To deal with the multi-modality of the KOH-IMSPE surface, our derivitive-based optimizations are wrapped in a multi-start scheme initialized at the best inputs found on a discrete LHS
candidate set. 

 \citet{leatherman2017designing} noted \textit{numerical
non-invertibility} of the covariance matrix in the implementation of their
minimum IMSPE joint acquisition of field and simulator data.  Their solution
involved only considering candidates spaced at a minimum distance from
existing design locations.  We noted similar instabilities when entertaining
close acquisitions.  However, we take a different approach to the problem by
carefully inverting matrices.  For example, when evaluating a quadratic form
like \texttt{t(k) \%*\% solve(K) \%*\% k},  for square matrix {\tt K} and column 
vector {\tt k}, we instead perform \texttt{t(k)
\%*\% solve(K,k)}.  This choice results in more calls to \texttt{solve()}, but
requires solving fewer terms for each call.  We ensure symmetry of inverted matrices
averaging the transpose of the inverse with itself: 
$\mathbf{K}^{-1} \leftarrow (\mathbf{K}^{-1} + (\mathbf{K}^{-1})^\top)/2$, for arbitrary square matrix \(\bm{K}\).

The remainder of this section describes two synthetic Monte Carlo (MC)
benchmarking exercises where actively learned KOH-IMSPE designs are compared
to the following space-filling alternatives: 
\begin{itemize}
\item[1.] LHS 
\item[2.] Uniformly random
\item[3.] MaxPro \citep{joseph2015maximum, MaxPro}
\item[4.] Sequential IMSPE on the computer model inputs
only, independent of any field data or calibration procedure.
\end{itemize}
We call this
comparator M-IMSPE, and it is calculated via \texttt{hetGP::IMSPE\_optim(...,\
h=0)\$par}.  
In addition to space filling designs, we also
include two variations of M-IMSPE which utilize information in
\(\bm{X}_{N_F}\) and \(\hat{\bm{u}}\).  
\begin{itemize}
\item[5.] The first one finds the minimum
M-IMSPE value in \(\bm{U}\)-space with the constraint that \(\tilde{\bm{x}}\)
must be one of the unique locations in \(\bm{X}_{N_F}\).  
\item[6.] The second minimizes
M-IMSPE in \(\bm{X}_N\)-space constrained by \(\tilde{\bm{u}} = \hat{\bm{u}}\).
We refer to these criteria as M-IMSPE \(\mid \tilde{\bm{x}} \in \bm{X}_{N_F}\)
and M-IMSPE \(\mid\tilde{\bm{u}} = \hat{\bm{u}}\) respectively.  
\end{itemize}
Comparisons
of KOH-IMSPE, LHS, M-IMSPE, and M-IMSPE \(\mid\tilde{\bm{u}} = \hat{\bm{u}}\)
are provided below, on several examples, while the remainder are shown in 
Supplement \ref{additional-experiments} to reduce clutter.  While
\textit{active learning} is not applicable via criteria which are not
model-based (LHS, random uniform, and MaxPro), we record the results from
these ``model-independent'' procedures in a sequential fashion in order to
draw comparisons across all options as design size increases.

Everything is re-randomized for each MC iteration, modulo each
 comparator design method, enumerated above, being
seeded with the same initial design: a
\(N_{M_0}\)-sized random subset of the full LHS design. To control variability, identical (but
 again random to each MC iteration) field data is shared between
each method. After each active learning acquisition GP hyperparameters and
\(\hat{\bm{U}}\) were updated. A natural metric for comparing these designs is
\begin{equation}\label{eq:rmse}
\mathrm{RMSE}
=
\sqrt{\frac{1}{N}\sum^{N}_{i = 1} (\mu(\bm{x}_i) - y_i)^2}
\end{equation}
calculated on the
predictions \(\mu(\bm{x}_i)\) from each method against a hold-out set of
(de-noised) field data outputs \(y_1, \dots, y_N\). These RMSE values are
saved after each acquisition, or as each additional element of a model-independent design is incorporated. We explore how the
distribution of these RMSE values changes as budgets \(N_M\) are increased
from \(N_{M_0}\) up to a final stopping budget determined \emph{ex-post} after
the best of the method(s) seem to have converged.

\hypertarget{sinexample}{%
\subsection{Sinusoid}\label{sinexample}}

Our first benchmarking example was used as the basis of
our running illustration, with details provided in Section \ref{illustration}. We entertained
1000 MC repetitions using \(N_{M_0} = 10\) initial space-filling design points and a final budget of \(N_M = 50\). Priors on GP hyperparameters were: \(p(\theta^M) = \mathrm{Gamma}(3/2, 2)\), \(p(\theta^B) = \mathrm{Gamma}(3/2, 5)\),
\(p(g_B) = \mathrm{Gamma}(3/2,7)\). For each MC iteration a
fixed, evenly spaced grid of 10 field data
points was evaluated twice, providing \(N_F = 20\) unique (randomly generated) values. RMSE for field predictions was computed using a \(N = \)100-point LHS test set, common to each method but novel to each MC iteration.

\begin{figure}[ht!]
{\centering
\includegraphics{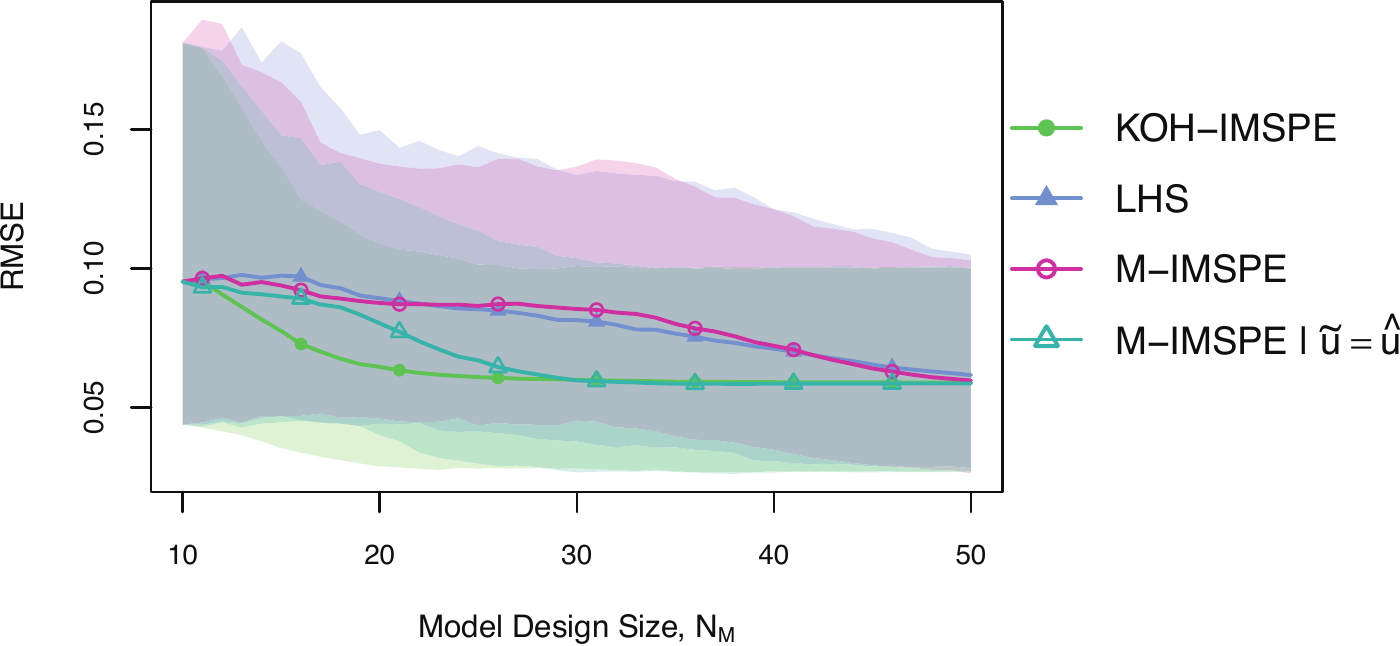} 
} \caption{Left: mean RMSE and 90\% quantiles for sinusoid data generating mechanism using KOH-IMSPE, LHS, M-IMSPE, and M-IMSPE \(\mid\tilde{\bm{u}} = \hat{\bm{u}}\) designs.}

\label{fig:sinusoidsrmse}
\end{figure}

Figure \ref{fig:sinusoidsrmse} summarizes the results of this experiment. The plot shows mean RMSE and 90\% quantiles over \(N_M\). There is high
variability in the results which may be attributed to the low signal/noise
ratio in the field data relative to computer model runs. Even so, after the
addition of just a few points, KOH-IMSPE clearly outperforms its competitors
on average. To control clutter in our visuals, competitors MaxPro, random
uniform, and M-IMSPE \(\mid \tilde{x} \in \bm{X}_{N_{F}}\), are relegated in Figure
\ref{fig:sinusoidsrmseappendix} provided with Supplement
\ref{additional-methods-for-sinusoid}.

Interestingly, with this example, there is little difference in performance
between the model-based default M-IMSPE and the model
independent space-filling designs. M-IMSPE \(\mid\tilde{\bm{u}} =
\hat{\bm{u}}\) shows better performance, through the mean and lower quantile
values, than the other comparators, but still clearly lags behind KOH-IMSPE.
Focusing on improving computer model prediction accuracy does not always
translate into improved field predictions, and a criterion that
does not allow exploration of the computer model away from \(\hat{\bm{u}}\)
may be too rigid. Estimation risk is the culprit here.
Relying on quality estimators of variance in low-signal settings to fine-tune
space-fillingness in a design is risky compared to targeting space-fillingness
geometrically. In higher-dimensional settings (as we shall see momentarily),
where notions of space-filling by variance are more nuanced, there is more
scope for improvement.

 In Supplement \ref{confounding}, we consider variations on this experiment
where we vary the the amplitude of the bias.  This serves two purposes.  One
is to explore estimation risk further: increasing the bias makes it hard to
separate signal from noise when relating the computer model to the field data.
The second is to explore how confounding may adversely affect our active
learning contribution.  We relegated these experiments to the Supplement
because, to our delight, they did not reveal any novel insights: the results
are basically the same as in Figure \ref{fig:sinusoidsrmse}.  When the bias is
higher the problem is harder for all design/modeling methods, but the relative
pecking order of those methods is the same.  Our active learning methods still
win. Although confounding presents an identifiability hazard when attempting
to estimate, and ascribe meaning to, a calibration parameter $\mathbf{u}$, we
have not seen any evidence that it prevents you from getting good predictions,
or as good as possible given a limited simulation budget.

Before we move on, we would like to quickly revisit Figure
\ref{fig:predvarfig}.  Comparing the top two panels, observe how KOH modeling
reduces the average predictive variance, but only slightly, in part because
the predictive variance given \(\bm{Y}_{N_F}\) alone is already low.  The
field data design is different for the experiment in this section, but the
principal still remains.  KOH modeling reduces predictive variance, but the
magnitude of the decrease is subject to a wide variety of effects including
data design, data size relative to input dimension, field data noise, initial
experimental design, and the utility of the simulator. While KOH-IMSPE was
clearly the best performing methodology we compared, the results are not
astounding, in part because the utility of KOH-IMSPE for a particular setting
is limited by the utility of KOH.  In the next example, there is more room for
improvement through KOH modeling and therefore KOH-IMSPE.

\hypertarget{gohbastos-problem}{%
\subsection{Goh/Bastos problem}\label{gohbastos-problem}}

Our second example comes from \citet{goh2013prediction}, which is adapted from \citet{bastos2009diagnostics}, but our treatment most closely resembles the slightly simpler setup described in Exercise 2 from Section 8 of \citet{gramacy2020surrogates}.
It involves a 2d design space (\(\bm{x}\)) and 2d calibration parameter (\(\bm{u}\)).
The data-generating mechanism is described as follows, where \(\bm{u}^\star = (0.2, 0.1)\).
\begin{align*}
& y^M(\bm{x},\bm{u}) =\\
& \quad  \left(1-\exp\left(-\frac{1}{2x_2}\right)\right)\frac{1000 u_1 x_1^3 + 1900 x_1^2 + 2092 x_1 + 60}{100 u_2 x_1^3 + 500 x_1^2 + 4x_1 + 20} \\
& b(\bm{x}) = \frac{10x_1^2 + 4x_2^2}{50x_1x_2 + 10}  \\
& Y^F(\bm{x}) = y^M \left( \bm{x}, \bm{u}^\star = \lbrack 0.2, 0.1 \rbrack\right) + b(\bm{x}) + \epsilon \\
& \epsilon \stackrel{\mathrm{iid}}{\sim} \mathcal{N}(0, 0.25^2)
\end{align*}

We performed 100 MC repetitions with  initial \(N_{M_0}\) = 30
and final
\(N_M = 130\). Field data was observed at 25 unique locations on an evenly
spaced grid in \(X\in\lbrack 0 , 1\rbrack ^2\), with two replicates at each
for a total of \(N_F = 50\). Priors for hyperparameters were as follows:
\(p(\theta^M) = \mathrm{Gamma}(3/2, 5/4)\),
\(p(\theta^B) = \mathrm{Gamma}(3/2, 5/2)\),
\(p(g_B) = \mathrm{Gamma}(3/2,1/20)\). RMSEs were calculated
on an out-of-sample (noise-free) testing set  calculated on
novel \(N = \) 1000-sized LHSs for each repetition.

\begin{figure}[ht!]
{\centering
\includegraphics{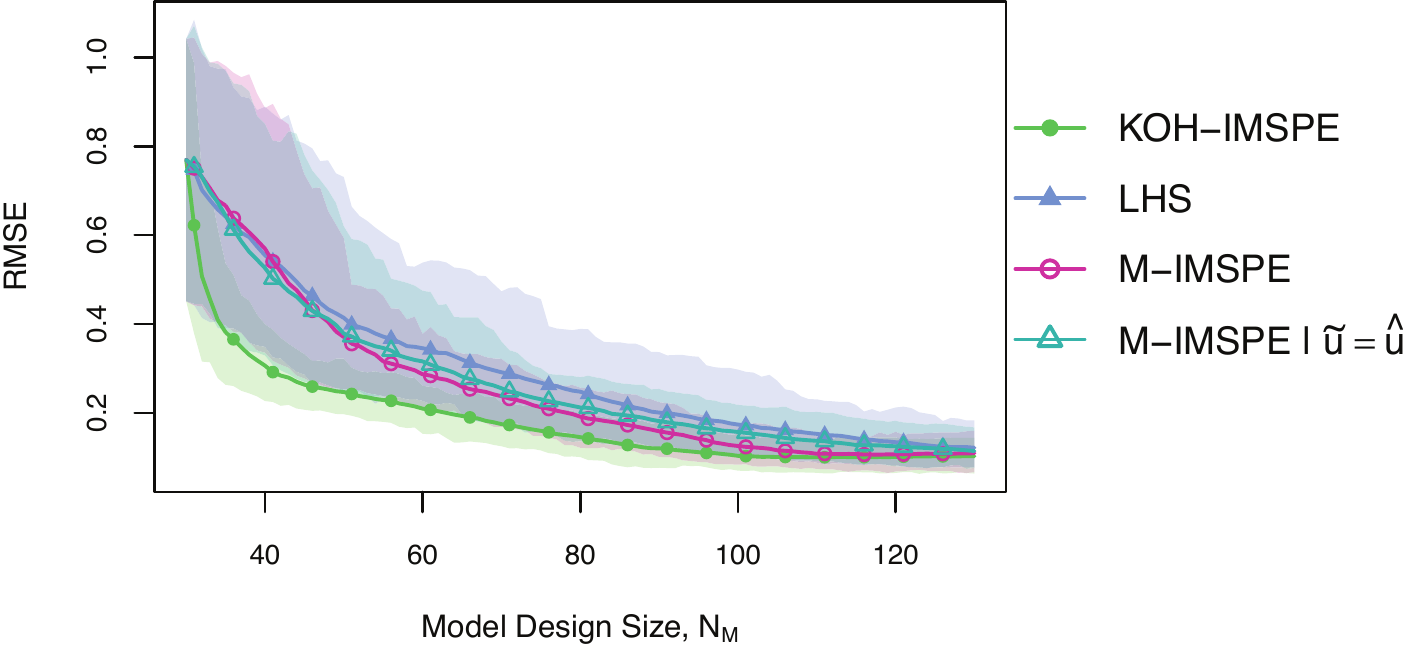} 
}
\caption{Left: mean RMSE and 90\% quantiles for data generating mechanism from
\citet{gramacy2020surrogates} (Ch 8, Ex 2) using KOH-IMSPE, LHS, M-IMSPE, and
M-IMSPE \(\mid\tilde{\bm{u}} = \hat{\bm{u}}\) designs.}\label{fig:surrogatesrmse}
\end{figure}

Figure \ref{fig:surrogatesrmse} shows the results in the same layout as Figure
\ref{fig:sinusoidsrmse}. As in that experiment, the remaining
comparators are shown in Figure \ref{fig:surrogatesrmseappendix},
Supplement \ref{additional-designs-for-gohbastos-problem}.
Observe that from \(N_M = 31\) to around \(N_M = 40\) there is little
discernible difference in RMSE between LHS, M-IMSPE, and M-IMSPE
\(\mid\tilde{\bm{u}} = \hat{\bm{u}}\) designs. However, mean RMSE for
KOH-IMSPE quickly dominates, and bounds for its 90\% quantile are much
narrower throughout the experiment.  Shortly after \(N_M = 40\) M-IMSPE shows consistent improvement over the
randomized space-filling methods, but M-IMSPE requires the acquisition of
another 70 data points before the method is competitive with KOH-IMSPE.
Due to the particulars of this problem, mean RMSE shows that
M-IMSPE \(\mid\tilde{\bm{u}} = \hat{\bm{u}}\) performs slightly worse than
default M-IMSPE as \(N_M\) increases, and the mean performance of M-IMSPE
\(\mid\tilde{\bm{x}} \in \bm{X}_{N_F}\) (shown in supplementary Figure
\ref{fig:surrogatesrmseappendix})  fluctuates with varying \(N_M\) relative to
others.  The details of a direct comparison between the M-IMSPE designs
differs somewhat from the previous example.  However, clearly the relative
flexibility and clear objective of KOH-IMSPE model-based active learning is
advantageous in this example, with KOH-IMSPE quickly providing the largest
benefit for improving accuracy of field predictions.

\hypertarget{reeapp}{%
\section{Solvent extraction of rare Earth elements}\label{reeapp}}

This work was motivated by an industrial application involving a chemical
process for concentrating Rare Earth Elements (REE), a significant portion of
which are allocated to \emph{high growth} green technologies, such as battery
alloys \citep{Goonan2011, balaram2019rare}. REEs include elements from the
lanthanide series, Yttrium, and Scandium \citep{VanGosen2014}. Liquid-liquid
extraction, also known as solvent extraction (SX), processes are often used to
concentrate rare earth elements \citep{gupta} from natural and recycled
sources. SX leverages the differing solubilities of various elements in
organic (oil) and aqueous (water) solutions to separate elemental products.

Testing SX plants is expensive due to the time required for the process to
reach steady state, and the difficulty of directly manipulating experimental
conditions. Active learning in the ``field'' is infeasible. Gathering data on
elemental concentrations across the organic and aqueous phases
passively through observation is much easier. SX chemical
reactions are governed by unknown chemical equilibrium constants, but
knowledge of exact values in this application is not important. Yet accurate
prediction of elemental equilibrium concentrations is imperative for technical
and economic analysis. Prediction of SX equilibria can benefit from the
additional information provided from a simulator via KOH. However, the high
dimensionality of the simulator parameter space and the requisite solutions of
systems of differential equations prohibits exhaustive evaluation. Active
learning here is essential, for developing an efficient simulation campaign.

 Here we provide the results of testing the predictive accuracy
of a KOH-IMSPE design for a REE experiment.
We have a small, \(N_F=\)
27-sized field data set. For these laboratory tests, three quantities
\(\bm{x}\) (coded to \([0,1]^3\)) were varied: the ratio of organic to aqueous
liquids, and the mols and volume of sodium hydroxide (NaOH) solution used to
adjust solution pH. We focus here on modeling the concentration of lanthanum
(La) in the water-based aqueous phase, our output \(y\)-variable. Simulation
of this quantity requires four (additional) chemical kinetic constants,
calibration parameters \(\bm{u}\) (coded to \([0,1]^4\)). Surrogate modeling
therefore requires the exploration of a 7d \((\bm{x},\bm{u})\)-space. Each run
outputs elemental concentrations after approximating the solution of a set of
differential equations through a Runge-Kutta routine. The simulator was run at
a low fidelity due to the intensive nature of the MC experiment, resulting in
a computation time of around 3.65 seconds on an 8-core Apple M1 Pro leveraging
Apple vecLib. KOH-IMSPE acquisition was fast even in comparison to a low fidelity simulator, requiring only 1.7 seconds on average to solve the 7d optimization problem.  Data from the MC experiment was obtained from multiple computers
and had total processor time of approximately 3.5 weeks.  Further technical
details can be found in Supplement \ref{sxappendix}; our \textsf{R}
implementation may be found in our repository along with other materials to
reproduce these experiments.

To manage expectations, we remark that \(N_F = 27\) is very small in a
seven-dimensional space. This has three consequences: (1) predictions on
these field data lean heavily on computer model simulations; (2)
information about promising \(\bm{u}\)-values is weak; (3) scope for out-of-sample
assessment of accuracy is limited and will have high MC error. Any design
which is space-filling in \(\bm{u}\) coordinates will perform about as well as
expected, leaving little scope for improvement with fancier alternatives, like
KOH-IMSPE. Nevertheless, we argue that a KOH-IMSPE active learning
strategy is worthwhile.

In each trial of our MC experiment, set up similarly to those in Section \ref{implementation}, we randomly held out 5 field data runs for out-of-sample RMSE assessments, so actually we used \(N_F=\) 22. We entertained an initial simulation design of size \(N_{M_0}=\) 50, and
active learning up to \(N_M=\) 300. Independent priors were as follows: \(p(\theta^M) = p(\theta^B) = \mathrm{Gamma}\left(3/2, 9/10\right), p(g) = \mathrm{Gamma}\left(3/2,1/20\right)\), and a \(\mathrm{Beta}\left(2,2\right)\) on each coordinate of \(\bm{u}\). We performed a total of 500 MC trials, each with a unique train-test partition and initial LHS.

\begin{figure}[ht!]
\includegraphics{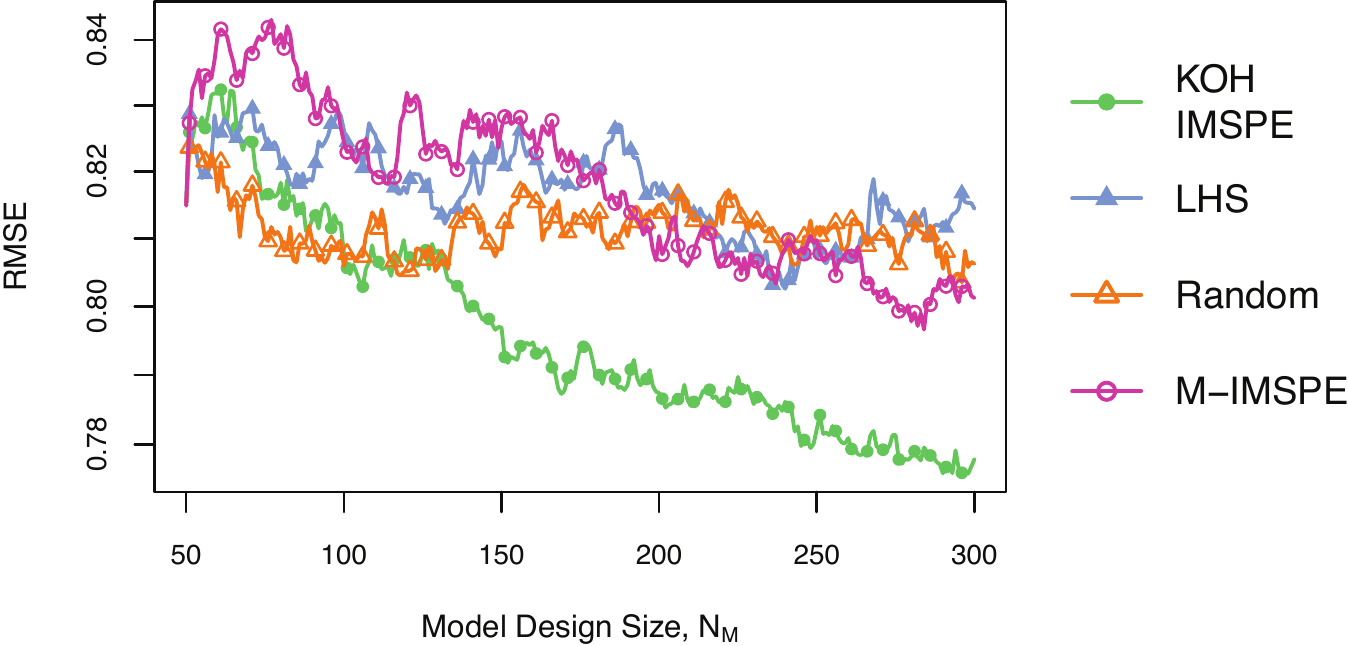} \caption{Left: mean RMSE calculated on a hold-out set for the SX application using KOH-IMSPE, LHS, random uniform, and M-IMSPE designs.}\label{fig:sxrmse}
\end{figure}

A plot of the average RMSE on the hold-out set is shown in Figure \ref{fig:sxrmse}. Error bars are removed to reduce clutter. Rather, we
report that the signal-to-noise is low on these RMSE values. Nevertheless, it is plain to see that, on
average, KOH-IMSPE outperforms its comparators.
At \(N_M = 300\) a one-sided paired Wilcoxon test of KOH-IMSPE versus M-IMSPE
(the second-best by mean) rejects the null by most conventional levels with a
\(p\)-value of \ensuremath{5.607\times 10^{-5}}. The signed rank
sum is \(T = -24,972\) using all \(n = 500\) MC experiments, and an effect
size of \(\left(\frac{|Z=3.9|}{\sqrt{n}}\right) = 0.17\).

\begin{figure}[ht!]
\includegraphics{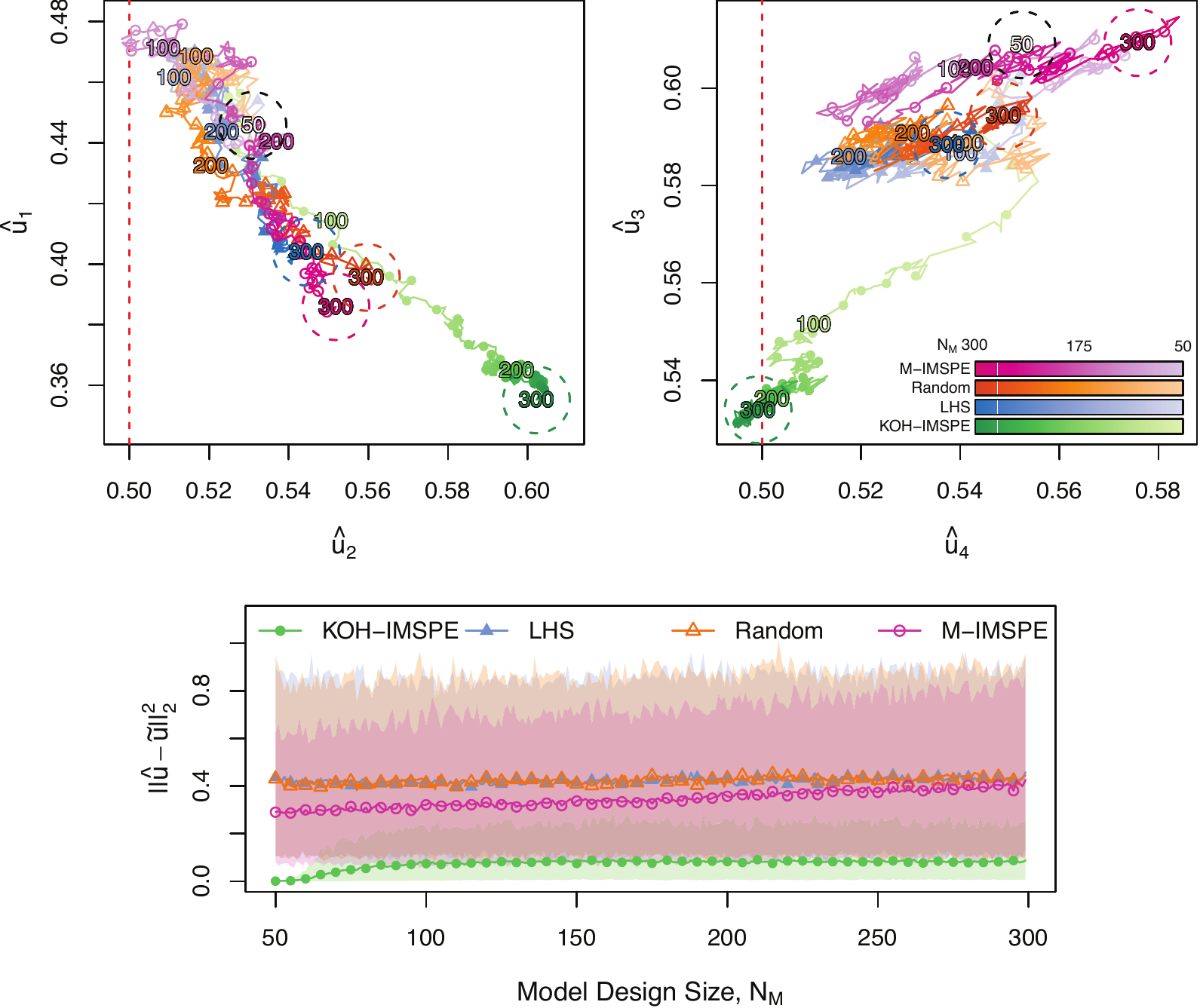} \caption{\textit{top panels:}  Mean path to convergence of $\hat{\bm{u}}$ for various designs used in the SX application. Initial and final estimates are circled.  Prior modes are shown as a red dashed line. \textit{bottom panel:}  Mean and 90\% quantiles of squared distance between $\hat{\bm{u}}$ and $\tilde{\bm{u}} \mid \hat{\bm{u}}$ for increasing $N_M$.}\label{fig:sxuhatconv}
\end{figure}

To further convince ourselves that our KOH-IMSPE designs are doing something interesting, we produced the plots in Figure \ref{fig:sxuhatconv}. These are inspired by Figure \ref{fig:twodfig}, which involved a more straightforward setup with one-dimensional \(\bm{u}\). Visuals in 4d are more complex. The top panel of Figure \ref{fig:sxuhatconv} shows the mean path of \(\hat{\bm{u}}\) as \(N_M\) increases. For every design type, the path line starts as an off-white color when \(N_M = 50\), and becomes darker as \(N_M\) approaches 300. Observe that the estimates obtained by the space-filling designs stay near, or take their time venturing away from the initial \(\hat{\bm{u}}\) obtained at \(N_M = 50\). By contrast, KOH-IMSPE quickly converges on a different subset of the space. Computer model acquisitions focused on improving prediction accuracy in the field improves convergence on \(\hat{\bm{u}}\), which helps reduce RMSE {[}Figure \ref{fig:sxrmse}{]}.

The bottom panel of Figure \ref{fig:sxuhatconv} provides some insight as to why KOH-IMSPE acquisition leads to more efficient \(\hat{\bm{u}}\) convergence. This plot provides mean squared distances and 90\% quantiles between \(\hat{\bm{u}}\) and the subsequent acquisition \(\tilde{\bm{u}}\). It is worth remarking that that LHS, Random, and M-IMSPE do not utilize \(\hat{\bm{u}}\) to find \(\tilde{\bm{u}}\). Therefore, for these space-filling design types \(||\tilde{\bm{u}} - \hat{\bm{u}}||^2_2\) is primarily related to how far each element of \(\hat{\bm{u}}\) is from 0.5. Observe that KOH-IMSPE, on the other hand, first collects points that are quite close to \(\hat{\bm{u}}\), with little variation. Then, as \(N_M\) increases, GP hyperparameters converge, and RMSE decreases, the variation in \(||\tilde{\bm{u}} - \hat{\bm{u}}||^2_2\) increases. We attribute this exploratory behavior to weight of the non-zero vectors in Eq. \eqref{eq:dkdutilde}. These increase -- become more important -- with increasing \(N_M\), pulling the location of the modes of KOH-IMSPE away from \(\hat{\bm{u}}\). Such exploratory behavior helps guard against pathologically bad estimates of \(\hat{\bm{u}}\), avoiding a vicious cycles that can plague active learning endeavors, especially in low-signal settings.

\hypertarget{discussec}{%
\section{Discussion}\label{discussec}}

Active learning is a weapon in the computer simulation experiment arsenal.
When computers are involved, the scope for human-free automation is higher
than in other settings where experimentation is required. Design strategies
governed by an active learning criterion have the potential to
reduce data related expenses in the pursuit of achieving a specific goal.
Here, we focus on a criterion which targets the goal of reduced
prediction error when an expensive uncalibrated simulator is available.  Our
KOH-IMSPE active leaning approach is tailor-made for the careful probing of
the (hyper)volume of the combined prediction and calibration inputs of a
simulator which takes hours, days, or even weeks to run.  Intuitively,
additional field data has been shown to be more desirable for improving field
prediction \citep{ranjan2011follow}.  However, in cases where additional field
data is impossible or difficult to gather, KOH modeling is a good, or possibly
the only, option for improving prediction accuracy.  In that setting, 
KOH-IMSPE is the active learning counterpart for efficient
collection of simulator data.

There are many factors in play for a KOH model including field
data size, field data locations, field data replicates, the magnitude of the
true bias, and the level of the true noise.  For settings where the functional
variance of field predictions is low, such as for a large and well spaced
\(\bm{X}_{N_F}\), KOH has little to offer, no matter the methodology used to
acquire simulator data.  Unfortunately, our motivating example in Section
\ref{reeapp} saw little benefit from KOH modeling. Marginalizing over all the
design methodologies, the results in Section \ref{reeapp} only show a slight
improvement in RMSE for increasing \(N_M\) on average.  We speculate that the
low utility of KOH in this setting is a product of small \(N_F\) relative
to the input dimension, and the difficulty in identifying noise from the true
bias with such a small \(N_F\) and no replicates.  In this setting, pathologies of
the specific problem limit the potential for reduction in RMSE via KOH, but
KOH-IMSPE still makes the most of this potential by performing ever so
slightly better than the comparators on average. 

 For settings where simulator data has a large potential for
reducing predictive variance, KOH-IMSPE benefits from this potential and
breaks away from the pack in terms of performance, analogous to the clear
results related to Section \ref{sinexample}.  The collection of empirical
results we present show that KOH-IMSPE performs just as well or better than
other designs suggested in the literature, with the magnitude of the benefit
of using KOH-IMSPE dependent on the utility of KOH.  The potential of KOH is
often not known before implementation.  Therefore, given the empirical
evidence provided, we suggest collecting expensive simulator data for KOH
modeling via KOH-IMSPE when the goal is the largest expected reduction in RMSE
for a given computational budget.

There are, of course, other design goals that might be of interest, i.e.,
beyond reduced predictive variance. We remarked in Section
\ref{introduction} that identifiability/confounding is a concern with KOH.
 Although a legitimate concern in many contexts, we have not
found that it affects the ability to obtain good designs for prediction, which
has been the main target of this research.  Our experiments, specifically in
Supplement \ref{confounding}, suggest that the entire modeling enterprise is
harder the more bias there is in the data-generating mechanism, but the best
designs for prediction still come from minimizing predictive variance (i.e.,
IMSPE).  We see no reason why the same underlying idea could not be ported to
fancier methods that attempt to mitigate identifiability issues
\citep[e.g.][]{plumlee2017bayesian,gu2018scaled}.  However, one would need to
work around the Monte Carlo being utilized in such contexts, which would pose
problems for derivative-based optimization of acquisition criteria.  Perhaps
candidate schemes \citep{gramacy2022triangulation} could help here. Other
potential add-ons might include batch acquisition for distributed simulation
environments.

It is natural to wonder if similar active learning principles could fruitfully
be deployed to augment field data campaigns. Our initial experiments with this
setup, admittedly using a cruder numerical IMSPE calculation, suggested
that there may not be not much difference between optimal
KOH-IMSPE and Field-IMSPE designs. Both methodologies space fill in \(\mathcal{X}\), but a thorough investigation could uncover some nuanced differences.  Application dependent, there may be less scope for human-free automation
in the design of field experiments in the KOH setting in comparison to automated simulator evaluation. However, an interesting
twist to batch KOH-IMSPE would be the joint augmentation of both field and
simulator data, for cases when more simulations can be run concurrent to the
collection of additional field data. In this setup there may be further scope
for automating aspects of design.


In our motivating SX example, we had some difficulties evaluating the gradient of KOH-IMSPE in a numerically stable fashion. We were able to work around these with careful engineering, but these ``hacks'' may compromise the portability of our subroutines. Using better conditioned Matérn covariance functions \citep{stein1999interpolation} may provide more stable gradient evaluations.
Bayesian inference of \(\hat{\bm{u}}\) coupled with an efficient evaluation of KOH-IMSPE criterion may provide improvements due to to further uncertainty quantification of the calibration parameter in small field data settings. Additionally, Bayesian optimization \citep{jones1998efficient} could be used to maximize the likelihood of \(\hat{\bm{u}}\) while probing the simulator input space, possibly allowing for reduced data requirements for convergence on \(\hat{\bm{u}}\). For simulators with multiple outputs a KOH-IMSPE criterion used with a cokriging model \citep{ver1998constructing} may be able to leverage larger amounts of information at each point for improved convergence of the calibration parameter.

\hypertarget{acknowledgements}{%
\subsection*{Acknowledgements}\label{acknowledgements}}
\addcontentsline{toc}{subsection}{Acknowledgements}

This manuscript has been authored with number LA-UR-24-23370 by Triad National Security under Contract with the U.S. Department of Energy, Office of Defense Nuclear Nonproliferation Research and Development. This research was funded by the National Nuclear Security Administration, Defense Nuclear Nonproliferation Research and Development (NNSA DNN R\&D). The authors acknowledge important interdisciplinary collaboration with scientists and engineers from LANL, LLNL, MSTS, PNNL, and SNL. Los Alamos National Laboratory is supported by the U.S. Department of Energy National Nuclear Security Administration under Contract No. DE-AC52-06NA25396. The United States Government retains and the publisher, by accepting the article for publication, acknowledges that the United States Government retains a non-exclusive, paid-up, irrevocable, world-wide license to publish or reproduce the published form of this manuscript, or allow others to do so, for United States Government purposes.

This work supported, in part, by the U.S. Department of Energy, Office of Science, Office of Advanced Scientific Computing Research and Office of High Energy Physics, Scientific Discovery through Advanced Computing (SciDAC) program under Award Number 0000231018; and by National Science Foundation award CMMI-2152679. Research presented in this article was also supported by the Laboratory Directed Research and Development program of Los Alamos National Laboratory under project number 20220188DR.  Los Alamos National Laboratory is operated by Triad National Security, LLC, for the National Nuclear Security Administration of U.S. Department of Energy (Contract No. 89233218CNA000001).

\hypertarget{supmat}{%
\subsection*{Supplementary Materials}\label{supmat}}
\addcontentsline{toc}{subsection}{Supplementary Materials}

Supplementary material contains details on the derivations in section A.  Section B provides details of the solvent extraction simulator.  Section C provides the results of additional experiments, with section C.1 and C.2 showing the results from additional experimental design comparators not included in Figure 4 and Figure 5 respectively.  Section C.3 provides details and results of a short experiment investigating confounding by examining RMSE and the distance between \(\hat{u}\) and \(u^{\star}\) when the magnitude of the bias function is varied.  The data generating mechanism in the beginning of Section 3.2 was used for section C.3.

\bibliography{pubdocs/bib}

\begin{thebibliography}{}

\bibitem[Arendt et~al., 2016]{arendt2016preposterior}
Arendt, P.~D., Apley, D.~W., and Chen, W. (2016).
\newblock A preposterior analysis to predict identifiability in the
  experimental calibration of computer models.
\newblock {\em IIE Transactions}, 48(1):75--88.

\bibitem[Ba and Joseph, 2018]{MaxPro}
Ba, S. and Joseph, V.~R. (2018).
\newblock {\em MaxPro: Maximum Projection Designs}.
\newblock R package version 4.1-2.

\bibitem[Balaram, 2019]{balaram2019rare}
Balaram, V. (2019).
\newblock Rare earth elements: A review of applications, occurrence,
  exploration, analysis, recycling, and environmental impact.
\newblock {\em Geoscience Frontiers}, 10(4):1285--1303.

\bibitem[Bastos and O’Hagan, 2009]{bastos2009diagnostics}
Bastos, L.~S. and O’Hagan, A. (2009).
\newblock Diagnostics for gaussian process emulators.
\newblock {\em Technometrics}, 51(4):425--438.

\bibitem[Bayarri et~al., 2009]{bayarri2009modularization}
Bayarri, M., Berger, J., and Liu, F. (2009).
\newblock Modularization in bayesian analysis, with emphasis on analysis of
  computer models.
\newblock {\em Bayesian Analysis}, 4(1):119--150.

\bibitem[Bayarri et~al., 2007]{bayarri2007framework}
Bayarri, M.~J., Berger, J.~O., Paulo, R., Sacks, J., Cafeo, J.~A., Cavendish,
  J., Lin, C.-H., and Tu, J. (2007).
\newblock A framework for validation of computer models.
\newblock {\em Technometrics}, 49(2):138--154.

\bibitem[Bernstein, 2009]{bernstein2009matrix}
Bernstein, D.~S. (2009).
\newblock {\em Matrix mathematics}.
\newblock Princeton university press.

\bibitem[Binois and Gramacy, 2021]{hetGP}
Binois, M. and Gramacy, R.~B. (2021).
\newblock {\em hetGP: Heteroskedastic Gaussian Process Modeling and Design
  under Replication}.
\newblock R package version 1.1.5.

\bibitem[Binois et~al., 2019]{binois2019replication}
Binois, M., Huang, J., Gramacy, R.~B., and Ludkovski, M. (2019).
\newblock Replication or exploration? sequential design for stochastic
  simulation experiments.
\newblock {\em Technometrics}, 61(1):7--23.

\bibitem[Brent, 2013]{brent2013algorithms}
Brent, R.~P. (2013).
\newblock {\em Algorithms for minimization without derivatives}.
\newblock Courier Corporation.

\bibitem[Brynjarsdottir and O'Hagan, 2014]{bryn2014learning}
Brynjarsdottir, J. and O'Hagan, A. (2014).
\newblock Learning about physical parameters: The importance of model
  discrepancy.
\newblock {\em Inverse Problems}, 30(11):114007.

\bibitem[Byrd et~al., 1995]{byrd1995limited}
Byrd, R., Qiu, P., Nocedal, J., , and Zhu, C. (1995).
\newblock A limited memory algorithm for bound constrained optimization.
\newblock {\em Journal on Scientific Computing}, 16(5):1190--1208.

\bibitem[Castillo et~al., 2019]{castillo2019bayesian}
Castillo, A.~R., Joseph, V.~R., and Kalidindi, S.~R. (2019).
\newblock Bayesian sequential design of experiments for extraction of
  single-crystal material properties from spherical indentation measurements on
  polycrystalline samples.
\newblock {\em JOM}, 71:2671--2679.

\bibitem[Chen et~al., 2022]{chen2022apik}
Chen, J., Chen, Z., Zhang, C., and Jeff~Wu, C. (2022).
\newblock Apik: Active physics-informed kriging model with partial differential
  equations.
\newblock {\em SIAM/ASA Journal on Uncertainty Quantification}, 10(1):481--506.

\bibitem[Cohn, 1994]{cohn1994neural}
Cohn, D. (1994).
\newblock Neural network exploration using optimal experiment design.
\newblock In {\em Advances in Neural Information Processing Systems}, pages
  679--686.

\bibitem[Cole et~al., 2021]{cole2021locally}
Cole, D.~A., Christianson, R.~B., and Gramacy, R.~B. (2021).
\newblock Locally induced gaussian processes for large-scale simulation
  experiments.
\newblock {\em Statistics and Computing}, 31(3):1--21.

\bibitem[Espenson, 1995]{espenson1995chemical}
Espenson, J.~H. (1995).
\newblock {\em Chemical kinetics and reaction mechanisms}, volume 102.
\newblock Citeseer.

\bibitem[Fer et~al., 2018]{fer2018linking}
Fer, I., Kelly, R., Moorcroft, P.~R., Richardson, A.~D., Cowdery, E.~M., and
  Dietze, M.~C. (2018).
\newblock Linking big models to big data: efficient ecosystem model calibration
  through bayesian model emulation.
\newblock {\em Biogeosciences}, 15(19):5801--5830.

\bibitem[Goh et~al., 2013]{goh2013prediction}
Goh, J., Bingham, D., Holloway, J.~P., Grosskopf, M.~J., Kuranz, C.~C., and
  Rutter, E. (2013).
\newblock Prediction and computer model calibration using outputs from
  multifidelity simulators.
\newblock {\em Technometrics}, 55(4):501--512.

\bibitem[Goonan, 2012]{Goonan2011}
Goonan, T.~G. (2012).
\newblock Rare earth elements-end use and recyclability.
\newblock In {\em Rare Earth Elements: Supply, Trade and Use Dynamics}, pages
  119--138.

\bibitem[Gramacy, 2016]{gramacy2016lagp}
Gramacy, R.~B. (2016).
\newblock lagp: large-scale spatial modeling via local approximate gaussian
  processes in r.
\newblock {\em Journal of Statistical Software}, 72:1--46.

\bibitem[Gramacy, 2020]{gramacy2020surrogates}
Gramacy, R.~B. (2020).
\newblock {\em Surrogates: {G}aussian Process Modeling, Design and Optimization
  for the Applied Sciences}.
\newblock Chapman Hall/CRC, Boca Raton, Florida.
\newblock \url{http://bobby.gramacy.com/surrogates/}.

\bibitem[Gramacy et~al., 2015]{gra:etal:2015}
Gramacy, R.~B., Bingham, D., Holloway, J.~P., Grosskopf, M.~J., Kuranz, C.~C.,
  Rutter, E., Trantham, M., and Drake, R.~P. (2015).
\newblock Calibrating a large computer experiment simulating radiative shock
  hydrodynamics.
\newblock {\em Annals of Applied Statistics}, 9(3):1141--1168.

\bibitem[Gramacy et~al., 2022]{gramacy2022triangulation}
Gramacy, R.~B., Sauer, A., and Wycoff, N. (2022).
\newblock Triangulation candidates for bayesian optimization.
\newblock In Oh, A.~H., Agarwal, A., Belgrave, D., and Cho, K., editors, {\em
  Advances in Neural Information Processing Systems}.

\bibitem[Gu, 2019]{gu2018jointly}
Gu, M. (2019).
\newblock Jointly robust prior for {G}aussian stochastic process in emulation,
  calibration and variable selection.
\newblock {\em Bayesian Analysis}, 14(3):857--885.

\bibitem[Gu and Wang, 2018]{gu2018scaled}
Gu, M. and Wang, L. (2018).
\newblock Scaled gaussian stochastic process for computer model calibration and
  prediction.
\newblock {\em SIAM/ASA Journal on Uncertainty Quantification},
  6(4):1555--1583.

\bibitem[Gupta and Krishnamurthy, 1992a]{gupta}
Gupta, C.~K. and Krishnamurthy, N. (1992a).
\newblock Extractive metallurgy of rare earths.
\newblock {\em International Materials Reviews}, 37(1):197--248.

\bibitem[Gupta and Krishnamurthy, 1992b]{gupta1992extractive}
Gupta, C.~K. and Krishnamurthy, N. (1992b).
\newblock Extractive metallurgy of rare earths.
\newblock {\em International materials reviews}, 37(1):197--248.

\bibitem[Higdon et~al., 2004]{Higdon:2004}
Higdon, D., Kennedy, M., Cavendish, J.~C., Cafeo, J.~A., and Ryne, R.~D.
  (2004).
\newblock Combining field data and computer simulations for calibration and
  prediction.
\newblock {\em SIAM Journal on Scientific Computing}, 26(2):448--466.

\bibitem[Huang et~al., 2020]{Huang:2018}
Huang, J., Gramacy, R.~B., Binois, M., and Libraschi, M. (2020).
\newblock On-site surrogates for large-scale calibration.
\newblock {\em Applied Stochastic Models in Business and Industry},
  36(2):283--304.
\newblock preprint on arXiv:1810.01903.

\bibitem[Johnson, 2008]{johnson:2008}
Johnson, L.~R. (2008).
\newblock Microcolony and biofilm formation as a survival strategy for
  bacteria.
\newblock {\em Journal of Theoretical Biology}, 251:24--34.

\bibitem[Johnson et~al., 1990]{johnson1990minimax}
Johnson, M.~E., Moore, L.~M., and Ylvisaker, D. (1990).
\newblock Minimax and maximin distance designs.
\newblock {\em Journal of statistical planning and inference}, 26(2):131--148.

\bibitem[Jones et~al., 1998]{jones1998efficient}
Jones, D., Schonlau, M., and Welch, W. (1998).
\newblock Efficient global optimization of expensive black-box functions.
\newblock {\em Journal of Global Optimization}, 13(4):455--492.

\bibitem[Joseph et~al., 2015]{joseph2015maximum}
Joseph, V.~R., Gul, E., and Ba, S. (2015).
\newblock Maximum projection designs for computer experiments.
\newblock {\em Biometrika}, 102(2):371--380.

\bibitem[Kennedy and O'Hagan, 2001]{kennedy2001bayesian}
Kennedy, M.~C. and O'Hagan, A. (2001).
\newblock Bayesian calibration of computer models.
\newblock {\em Journal of the Royal Statistical Society: Series B (Statistical
  Methodology)}, 63(3):425--464.

\bibitem[Krishna et~al., 2021]{krishna2021robust}
Krishna, A., Joseph, V.~R., Ba, S., Brenneman, W.~A., and Myers, W.~R. (2021).
\newblock Robust experimental designs for model calibration.
\newblock {\em Journal of Quality Technology}, pages 1--12.

\bibitem[Leatherman et~al., 2017]{leatherman2017designing}
Leatherman, E.~R., Dean, A.~M., and Santner, T.~J. (2017).
\newblock Designing combined physical and computer experiments to maximize
  prediction accuracy.
\newblock {\em Computational Statistics \& Data Analysis}, 113:346--362.

\bibitem[Leatherman et~al., 2018]{leatherman2018computer}
Leatherman, E.~R., Santner, T.~J., and Dean, A.~M. (2018).
\newblock Computer experiment designs for accurate prediction.
\newblock {\em Statistics and Computing}, 28:739--751.

\bibitem[MacKay, 1992]{mackay1992information}
MacKay, D.~J. (1992).
\newblock Information-based objective functions for active data selection.
\newblock {\em Neural computation}, 4(4):590--604.

\bibitem[Marrel et~al., 2009]{marrel2009calculations}
Marrel, A., Iooss, B., Laurent, B., and Roustant, O. (2009).
\newblock Calculations of {S}obol indices for the {G}aussian process metamodel.
\newblock {\em Reliability Engineering \& System Safety}, 94(3):742--751.

\bibitem[Matheron, 1963]{matheron1963principles}
Matheron, G. (1963).
\newblock Principles of geostatistics.
\newblock {\em Economic geology}, 58(8):1246--1266.

\bibitem[McKay et~al., 2000]{mckay2000comparison}
McKay, M.~D., Beckman, R.~J., and Conover, W.~J. (2000).
\newblock A comparison of three methods for selecting values of input variables
  in the analysis of output from a computer code.
\newblock {\em Technometrics}, 42(1):55--61.

\bibitem[Morris, 2015]{morris2015physical}
Morris, M.~D. (2015).
\newblock Physical experimental design in support of computer model
  development.
\newblock {\em Technometrics}, 57(1):45--53.

\bibitem[Nelder and Mead, 1965]{nelder1965simplex}
Nelder, J.~A. and Mead, R. (1965).
\newblock A simplex method for function minimization.
\newblock {\em The computer journal}, 7(4):308--313.

\bibitem[Plumlee, 2017]{plumlee2017bayesian}
Plumlee, M. (2017).
\newblock Bayesian calibration of inexact computer models.
\newblock {\em Journal of the American Statistical Association},
  112(519):1274--1285.

\bibitem[Plumlee, 2019]{plumlee2019}
Plumlee, M. (2019).
\newblock Computer model calibration with confidence and consistency.
\newblock {\em Journal of the Royal Statistical Society: Series B},
  81(3):519--545.

\bibitem[Ranjan et~al., 2011]{ranjan2011follow}
Ranjan, P., Lu, W., Bingham, D., Reese, S., Williams, B.~J., Chou, C.-C., Doss,
  F., Grosskopf, M., and Holloway, J.~P. (2011).
\newblock Follow-up experimental designs for computer models and physical
  processes.
\newblock {\em Journal of Statistical Theory and Practice}, 5(1):119--136.

\bibitem[Sacks et~al., 1989]{sacks1989design}
Sacks, J., Welch, W.~J., Mitchell, T.~J., and Wynn, H.~P. (1989).
\newblock Design and analysis of computer experiments.
\newblock {\em Statistical science}, 4(4):409--423.

\bibitem[Santner et~al., 2018]{santner2018design}
Santner, T., Williams, B., and Notz, W. (2018).
\newblock {\em The Design and Analysis of Computer Experiments, Second
  Edition}.
\newblock Springer--Verlag, New York, NY.

\bibitem[Sauer et~al., 2022]{sauer2021active}
Sauer, A., Gramacy, R.~B., and Higdon, D. (2022).
\newblock Active learning for deep gaussian process surrogates.
\newblock {\em Technometrics}, pages 1--15.

\bibitem[Seo et~al., 2000]{seo2000gaussian}
Seo, S., Wallat, M., Graepel, T., and Obermayer, K. (2000).
\newblock Gaussian process regression: Active data selection and test point
  rejection.
\newblock In {\em Mustererkennung 2000}, pages 27--34. Springer.

\bibitem[Stein, 1999]{stein1999interpolation}
Stein, M.~L. (1999).
\newblock {\em Interpolation of spatial data: some theory for kriging}.
\newblock Springer Science \& Business Media.

\bibitem[S{\"u}rer et~al., 2023]{surer2023sequential}
S{\"u}rer, {\"O}., Plumlee, M., and Wild, S.~M. (2023).
\newblock Sequential bayesian experimental design for calibration of expensive
  simulation models.
\newblock {\em arXiv preprint arXiv:2305.16506}.

\bibitem[Tuo and Wu, 2015]{tuo2015}
Tuo, R. and Wu, C. F.~J. (2015).
\newblock Efficient calibration for imperfect computer models.
\newblock {\em Annals of Statistics}, 43(6):2331--2352.

\bibitem[Tuo and Wu, 2016]{tuo2016}
Tuo, R. and Wu, C. F.~J. (2016).
\newblock A theoretical framework for calibration in computer models:
  Parameterization, estimation and convergence properties.
\newblock {\em Journal of Uncertainty Quantification}, 4:767--795.

\bibitem[Van~Gosen et~al., 2014]{VanGosen2014}
Van~Gosen, B.~S., Verplanck, P.~L., Long, K.~R., Gambogi, J., and Seal, R.~R.
  (2014).
\newblock The rare-earth elements; vital to modern technologies and lifestyles.
\newblock {\em Fact Sheet - U. S. Geological Survey}.

\bibitem[Ver~Hoef and Barry, 1998]{ver1998constructing}
Ver~Hoef, J.~M. and Barry, R.~P. (1998).
\newblock Constructing and fitting models for cokriging and multivariable
  spatial prediction.
\newblock {\em Journal of Statistical Planning and Inference}, 69(2):275--294.

\bibitem[Wei et~al., 2015]{wei2015submodularity}
Wei, K., Iyer, R., and Bilmes, J. (2015).
\newblock Submodularity in data subset selection and active learning.
\newblock In {\em International Conference on Machine Learning}, pages
  1954--1963. PMLR.

\bibitem[Williams et~al., 2011]{williams2011batch}
Williams, B.~J., Loeppky, J.~L., Moore, L.~M., and Macklem, M.~S. (2011).
\newblock Batch sequential design to achieve predictive maturity with
  calibrated computer models.
\newblock {\em Reliability Engineering \& System Safety}, 96(9):1208--1219.

\bibitem[Williams and Rasmussen, 2006]{williams2006gaussian}
Williams, C.~K. and Rasmussen, C.~E. (2006).
\newblock {\em Gaussian processes for machine learning}, volume~2.
\newblock MIT press Cambridge, MA.

\bibitem[Wong et~al., 2017]{wong2017}
Wong, R. K.~W., Storlie, C.~B., and Lee, T. C.~M. (2017).
\newblock A frequentist approach to computer model calibration.
\newblock {\em Journal of the Royal Statistical Society: Series B (Statistical
  Methodology)}, 79(2):635--648.

\bibitem[Wycoff et~al., 2021]{wycoff2021jcgs}
Wycoff, N., Binois, M., and Wild, S.~M. (2021).
\newblock Sequential learning of active subspaces.
\newblock {\em Journal of Computational and Graphical Statistics},
  30(4):1224--1237.

\end{thebibliography}

\hypertarget{appendix-appendix}{%
\appendix}

\pagebreak
\section*{Supplementary Material}

\hypertarget{kennedy-and-ohagan-imspe-derivations}{%
\section{Kennedy and O'Hagan IMSPE Derivations}\label{kennedy-and-ohagan-imspe-derivations}}

Here we provide the analytical results required for fast
KOH-IMSPE implementation using Gaussian covariance functions.  For an active
learning KOH-IMSPE design which minimizes the criterion:
\begin{equation*}
\mathrm{IMSPE}([\bm{X}_{N_M + 1}, \bm{U}_{N_M + 1}]) = \nu_M + \nu_B - \bm{1}^\top \left(\left[\bm{\Sigma}^{M,B}_{N_F + N_M + 1}\right]^{-1} \circ \left(\nu_M^2 \bm{W}^{M,M} +  2 \nu_M \nu_B \bm{W}^{M,B} +  \nu_B^2 \bm{W}^{B,B}\right)\right)\bm{1}.
\end{equation*}
Closed form solutions of the integrals with respect to the predictive location
\(\bm{x}\), \(\bm{W}^{M,M}, \bm{W}^{M,B}, \bm{W}^{B,B}\), and derivatives of
\(\bm{\Sigma}^{M,B}_{N_F + N_M + 1}, \bm{W}^{M,M}, \bm{W}^{M,B}\), and
\(\bm{W}^{B,B}\) with respect to the additional design location
\([\tilde{\bm{x}}_{1 \times p},\tilde{\bm{u}}_{1\times s}]_{1 \times d}\) are
provided.  All non-zero functions making up the elements of the matrices
\(\bm{W}^{M,M}, \bm{W}^{M,B}, \bm{W}^{B,B}\) are the result of integrating
over predictive \(\bm{x}\),  and only non-zero functions in the last row and
column of \(\bm{\Sigma}^{M,B}_{N_F + N_M + 1}, \bm{W}^{M,M}, \bm{W}^{M,B}\),
and \(\bm{W}^{B,B}\) contain \([\tilde{\bm{x}}, \tilde{\bm{u}}]\).  All
KOH-IMSPE and KOH-IMSPE gradient evaluations are conditioned on parameter
estimates of \(\hat{\bm{u}}_{1\times x}, \nu_M, \nu_B, \bm{\theta}^M_{1\times
p}, \bm{\theta}^B_{1\times s}, g\).

\hypertarget{intA}{%
\subsection{Integrals}\label{intA}}

Details are provided for evaluating the integrals required to calculate
KOH-IMSPE in closed form when the GPs used in modeling the computer
simulation and bias function both utilize a separable Gaussian covariance kernel. The provided derivations are for a uniformly rectangular \(\bm{X}_N\) space with inputs scaled to \([0,1]^p\) and a model conditioned on a point estimate of \(\hat{\bm{U}}\). The form for the integral is shown as \eqref{eq:int-wrt-x} multiplied by \eqref{eq:int-wrt-u}, and can also be noted as the Hadamard product \(\bm{W}^{\alpha, \beta}(\bm{X}_N) \circ \bm{W}^{\alpha, \beta}(\bm{U})\) Below, the variables \(\alpha,\beta\) express the combinations of the model and bias covariance kernels \((M,M); (M,B); (B,B)\), \(i,j\) indexes all data locations from 1 to \(N_F + N_M\), and \(\bm{x},\hat{\bm{u}}\) without an \(i,j\) subscript is the field data predictive location.
\begin{align}
w^{\alpha, \beta}_{i,j} &= \int_{[0,1]^p} k^{\alpha}([\bm{x}_i, \bm{u}_i], [\bm{x}, \hat{\bm{u}}])k^{\beta}([\bm{x}, \hat{\bm{u}}], [\bm{x}_j, \bm{u}_j])dx_1, \dots dx_p \notag \\
&= \prod_{l = 1}^p \int_0^1 \exp\left(-\frac{(x_{i,l} - x_l)^2}{\theta^\alpha_l}\right)\exp\left(-\frac{(x_{j,l} - x_l)^2}{\theta^\beta_l}\right)dx_l \label{eq:int-wrt-x}\\
&\qquad\quad\times\prod_{l = p + 1}^d \exp\left(-\frac{(u_{i,l} - \hat{u}_l)^2}{\theta^\alpha_l}\right)\exp\left(-\frac{(u_{j,l} - \hat{u}_l)^2}{\theta^\beta_l}\right) \label{eq:int-wrt-u}
\end{align}
To make notation simple and more compact, we do not state the bounds of integration going forward, and integrals should always be assumed to be evaluated from 0 to 1. This important note is not repeated again and again below in order to make the text more compact. For compactness we assume that \(N_M\) contains the point augmented to the collected data for a sequential design application in this section.

Derivatives are provided for a gradient based search of the additional computer model point \([\tilde{\bm{x}},\tilde{\bm{u}}]\) which minimizes KOH-IMSPE. The expressions provided must be multiplied by remaining \(d-1\) elements of the products \eqref{eq:int-wrt-x} and \eqref{eq:int-wrt-u}. For example, for \(l = 2 = p; d = 3\) find \(\frac{\partial \bm{W}^{M,M}}{\partial \tilde{x}_2} = \bm{W}^{M,M}(\bm{X}_1) \circ \frac{\partial \bm{W}^{M,M}(X_2)}{\partial \tilde{x}_2} \circ \bm{W}^{M,M}(U_1)\), while only the expression for \(\frac{\partial \bm{W}^{M,M}(X_2)}{\partial \tilde{x}_2}\) is provided below.

\hypertarget{wmm}{%
\subsubsection{\texorpdfstring{\(\bm{W}^{M,M}\)}{W\^{}\{M,M\}}}\label{wmm}}

The integral to be solved in order to find \(\bm{W}^{M,M}\):
\begin{equation}
\bm{W}^{M,M}  =  \begin{bmatrix}
\int \bm{k}([\bm{X}_{N_F}, \hat{\bm{U}}],[\bm{x}, \hat{\bm{u}}])\bm{k}([\bm{x}, \hat{\bm{u}}],[\bm{X}_{N_F},\hat{\bm{U}}]) d\bm{x} & \int \bm{k}([\bm{X}_{N_F}, \hat{\bm{U}}], [\bm{x}, \hat{\bm{u}}])\bm{k}([\bm{x},\hat{\bm{u}}], [\bm{X}_{N_M}, \bm{U}_{N_M}])  d\bm{x}\\
\int \bm{k}([\bm{X}_{N_M}, \bm{U}_{N_M}], [\bm{x}, \hat{\bm{u}}])\bm{k}([\bm{x},\hat{\bm{u}}], [\bm{X}_{N_F}, \hat{\bm{U}}]) d\bm{x} & \int \bm{k}([\bm{X}_{N_M}, \bm{U}_{N_M}], [\bm{x},\hat{\bm{u}}])\bm{k}([\bm{x}, \hat{\bm{u}}], [\bm{X}_{N_M}, \bm{U}_{N_M}]) d\bm{x}
\end{bmatrix} \label{eq:wmmblockmat}
\end{equation}
The \((i,j)^\mathrm{th}\) element of \(w^{M,M}\) corresponding to design location \(\bm{x}\) and \eqref{eq:int-wrt-x} can be found as:
\begin{equation}
w^{M,M}_{i,j}(\bm{x}) = \prod_{l = 1}^p \frac{\sqrt{2\pi \theta^M_l}}{4}\exp\left(- \frac{(x_{i,l} - x_{j,l})^2}{2\theta_l^M} \right)\left(\mathrm{erf}\left(\frac{2-(x_{i,l} + x_{j,l})}{\sqrt{2\theta_l^M}}\right) + \mathrm{erf}\left(\frac{x_{i,l} + x_{j,l}}{\sqrt{2\theta_l^M}}\right)\right)
\label{eq:wmmint}
\end{equation}
Where \(\mathrm{erf}()\) is the Gauss error function. Elements of \(\bm{W}^{M,M}\) related to \(\bm{U}\) space and the product \eqref{eq:int-wrt-u} can be found as \eqref{eq:wmm-wrtu}, where \(\bm{W}^{M,M} = \bm{W}^{M,M}(\bm{X}_N) \circ \bm{W}^{M,M}(\bm{U})\), and \(\bm{J}_{N_F \times N_F}\) is a square matrix of ones with a dimension of \(N_F\)
\begin{equation}
\bm{W}^{M,M}(\bm{U}) = 
\begin{bmatrix}
\bm{J}_{N_F \times N_F} & \bm{1}_{N_F} \bm{k}(\hat{\bm{u}}, \bm{U}_{N_M})\\
\bm{k}(\bm{U}_{N_M}, \hat{\bm{u}}) \bm{1}^{\top}_{N_F} & \bm{k}(\bm{U}_{N_M}, \hat{\bm{u}})\bm{k}(\hat{\bm{u}}, \bm{U}_{N_M})
\end{bmatrix}
\label{eq:wmm-wrtu}
\end{equation}
Differentiation of \(\bm{W}^{M,M}\) is required to supply the optimization routine a gradient for minimization. Note, that only the last column and row of \(\frac{\partial \bm{W}^{M,M}}{\partial[\tilde{\bm{x}},\tilde{\bm{u}}]_l} \ \forall: l\) are non-zero. Differentiating with respect to an \(\tilde{x}_l\) changes the \(l^\mathrm{th}\) element of the product in \eqref{eq:wmmint} into the following:
\begin{equation}
\begin{split}
\frac{\partial w^{M,M}(x_{i,l}, \tilde{x}_l)}{\partial \tilde{x}_l} &= \sqrt{\frac{\pi}{2}}\exp\left(-\frac{(x_{i,l} - \tilde{x}_l)^2}{2\theta^M_l}\right)\left((x_{i,l} - \tilde{x}_l)\frac{ \mathrm{erf}\left( \frac{2-(x_{i,l}+\tilde{x}_l)}{\sqrt{2\theta^M_l}}\right) + \mathrm{erf}\left(\frac{x_{i,l}+\tilde{x}_l}{\sqrt{2\theta^M_l}}\right)}{2\sqrt{\theta^M_l}} + \right. \\ 
&\left. \phantom{\frac{ \mathrm{erf}\left( \frac{2-(x_{i,l}-\tilde{x}_l)}{\sqrt{2\theta^M_l}}\right)}{2\sqrt{\theta^M_l}}} \frac{1}{2}\sqrt{\frac{2}{\pi}}\left(\exp\left(-\frac{(x_{i,l}+\tilde{x}_l)^2}{2\theta^M_l}\right) - \exp\left(-\frac{(2-x_{i,l}-\tilde{x}_l)^2}{2\theta^M_l}\right)\right)\right)
\end{split}
\notag
\end{equation}
However, for the
case where \(x_{i,l} = \tilde{x}_l\) (the bottom right corner of \(\bm{W}^{M,M}\)), instead the derivative in
\eqref{eq:dwbbxtilxtil} should be used.
\begin{equation}
\frac{\partial w^{M,M}(\tilde{x}_l, \tilde{x}_l)}{\partial \tilde{x}_l} = \exp\left(-\frac{2\tilde{x}_l^2}{\theta^M_l}\right) - \exp\left(-\frac{2(\tilde{x}_l-1)^2}{\theta^M_l}\right)
\label{eq:dwbbxtilxtil}
\end{equation}
Differentiation with respect to \(\bm{U}\) space produces the following matrix, which similar to \eqref{eq:wmm-wrtu} should be element wise multiplied by the remaining functions evaluated to create \(\bm{W}^{M,M}\) related to the remaining \(d-1\) dimensions of the input space.
\begin{equation}
\frac{\partial \bm{W}^{M,M}(\bm{U})}{\partial \tilde{u}_l}=
\begin{bmatrix}
\bm{0}_{N_F \times N_F} & \bm{0}_{(N_F - 1) \times (N_M -1)} & \bm{1}_{N_F} \frac{\partial k(\hat{\bm{u}}, \tilde{\bm{u}})}{\partial \tilde{u}_l}\\
\bm{0}_{(N_M -1) \times N_F} & \bm{0}_{(N_M - 1)\times (N_M -1)} & \bm{k}(\bm{U}_{N_M -1}, \hat{\bm{u}})\frac{\partial k(\hat{\bm{u}}, \tilde{\bm{u}})}{\partial \tilde{u}_l}\\
\bm{1}_{N_F}^{\top} \frac{\partial k(\tilde{\bm{u}}, \hat{\bm{u}})}{\partial \tilde{u}_l} & \bm{k}(\hat{\bm{u}}, \bm{U}_{N_M -1})\frac{\partial k( \tilde{\bm{u}}, \hat{\bm{u}})}{\partial \tilde{u}_l} & \frac{\partial k(\tilde{\bm{u}}, \hat{\bm{u}})}{\partial \tilde{u}_l}k(\hat{\bm{u}}, \tilde{\bm{u}}) + k(\tilde{\bm{u}}, \hat{\bm{u}})\frac{\partial k(\hat{\bm{u}}, \tilde{\bm{u}})}{\partial \tilde{u}_l}
\end{bmatrix}
\label{eq:dwmm-du}
\end{equation}

\hypertarget{wmb}{%
\subsubsection{\texorpdfstring{\(\bm{W}^{M,B}\)}{W\^{}\{M,B\}}}\label{wmb}}

\begin{equation}
\bm{W}^{M,B}  =  \begin{bmatrix}
\int \bm{k}([\bm{X}_{N_F}, \hat{\bm{U}}], [\bm{x}, \hat{\bm{u}}])\bm{k}^B(\bm{x}, \bm{X}_{N_F})d\bm{x} & \bm{0}_{N_F \times N_M}\\
\int \bm{k}([\bm{X}_{N_M}, \bm{U}_{N_M}], [\bm{x}, \hat{\bm{u}}])\bm{k}^B(\bm{x}, \bm{X}_{N_F})d\bm{x} & \bm{0}_{N_M \times N_M}
\end{bmatrix}
\label{eq:wmbblockmat}
\end{equation}
Solving for \eqref{eq:wmbblockmat} requires particular attention to the fact that \(\bm{k}\) and \(\bm{k}^B\) have
different lengthscale values (\(\bm{\theta}^M\) and \(\bm{\theta}^B\)) for the same dimension. The result is shown below, where \(x_j\) is always taken to be to the field data and from the bias kernel.
\begin{equation}
\begin{split}
w^{M,B}_{i,j}(x) &= \prod_{l = 1}^p \exp \left(-\frac{(x_{j,l}-x_{i,l})^2}{\theta^B_l + \theta_l^M}\right)\left(\frac{1}{2}\sqrt{\pi\left(\frac{1}{\theta_l^M}+\frac{1}{\theta^B_l}\right)^{-1}}\right) \left(\mathrm{erf}\left(\frac{\left(\frac{\theta^B_l x_{i,l} + \theta_l^M x_{j,l}}{\theta^B_l + \theta_l^M}\right)}{\sqrt{\left(\frac{1}{\theta_l^M} + \frac{1}{\theta^B_l}\right)^{-1}}}\right)\right. \\
& \qquad - \left.\mathrm{erf}\left(\frac{\left(\frac{\theta^B_l x_{i,l} + \theta_l^M x_{j,l}}{\theta^B_l + \theta_l^M}\right)-1}{\sqrt{\left(\frac{1}{\theta_l^M} + \frac{1}{\theta^B_l}\right)^{-1}}}\right)\right)
\end{split}
\label{eq:wmbint}
\end{equation}
Functions related to \(\bm{U}\) space can be found as \eqref{eq:wmbu}, where \(\bm{W}^{M,B} = \bm{W}^{M,B}(\bm{X}_N) \circ \bm{W}^{M,B}(\bm{U}_{N_M})\), and \(\bm{J}_{N_F \times N_F}\) is a square matrix of ones with a dimension of \(N_F\),
\begin{equation}
\bm{W}^{M,B}(\bm{U})=
\begin{bmatrix}
\bm{J}_{N_F \times N_F} & \bm{0}_{N_F \times N_M}\\
\bm{k}(\bm{U}, \hat{\bm{u}})\bm{1}^{\top}_{N_F} & \bm{0}_{N_M \times N_M}
\end{bmatrix}
\label{eq:wmbu}
\end{equation}

Differentiation of \eqref{eq:wmbint} with respect to \(\tilde{x}\) for
gradient based minimization of KOH-IMSPE produces:
\begin{equation}
\begin{split}
&\frac{\partial w^{M,B}(\tilde{x}_l,x_{j,l})}{\partial \tilde{x}_l} = \\
&\frac{e^{-\frac{(\tilde{x}_l - x_{j,l})^2}{\theta^B_l + \theta_l^M}}}{\theta_l^M + \theta^B_l} \left(\sqrt{\pi \left(\frac{1}{\theta^B_l} + \frac{1}{\theta_l^M}\right)^{-1}}(x_{j,l} -\tilde{x}_l)\left(\mathrm{erf}\left(\frac{\left(\frac{\theta_l^M x_{j,l} + \theta^B_l \tilde{x}_l}{\theta_l^M + \theta^B_l}\right)}{\sqrt{\left(\frac{1}{\theta_l^M} + \frac{1}{\theta^B_l}\right)^{-1}}}\right) - \mathrm{erf}\left(\frac{\left(\frac{\theta_l^M x_{j,l} + \theta^B_l \tilde{x}_l}{\theta_l^M + \theta^B_l}\right)-1}{\sqrt{\left(\frac{1}{\theta_l^M} + \frac{1}{\theta^B_l}\right)^{-1}}}\right)\right) + \right. \\
& \left. \phantom{\mathrm{erf}\left(\frac{\left(\frac{\theta_l^M x_{j,l} + \theta^B_l \tilde{x}_l}{\theta_l^M + \theta^B_l}\right)}{\sqrt{\left(\frac{1}{\theta_l^M} + \frac{1}{\theta^B_l}\right)^{-1}}}\right)} \theta^B_l \left(e^{-\frac{\left(\frac{1}{\theta_l^M} + \frac{1}{\theta^B_l}\right)(\theta_l^M x_{j,l} + \theta^B_l \tilde{x}_l)^2}{(\theta_l^M + \theta^B_l)^2}} - e^{-\left(\frac{1}{\theta_l^M} + \frac{1}{\theta^B_l}\right)\left(\frac{\theta_l^M x_{j,l} + \theta^B_l \tilde{x}_l}{\theta_l^M + \theta^B_l} - 1\right)^2}\right)\right)
\end{split} \notag
\end{equation}
Once again, because of the form of \eqref{eq:wmbblockmat}, \(x_j\) only corresponds to field data from the bias covariance functions
\begin{equation}
\frac{\partial \bm{W}^{M,B}(\bm{U})}{\partial \tilde{u}_l}=
\begin{bmatrix}
\bm{0}_{N_F \times N_F} & \bm{0}_{N_F \times (N_M)}\\
\bm{0}_{(N_M - 1) \times N_F} & \bm{0}_{(N_M -1) \times N_M}\\
\frac{\partial k(\tilde{\bm{u}}, \hat{\bm{u}})}{\partial \tilde{u}_l} \bm{1}^{\top}_{N_F} & 0_{1 \times 1}
\end{bmatrix}
\notag
\end{equation}

\hypertarget{wbb}{%
\subsubsection{\texorpdfstring{\(\bm{W}^{B,B}\)}{W\^{}\{B,B\}}}\label{wbb}}

\begin{equation}
\bm{W}^{B,B}=
\begin{bmatrix}
\int \bm{k}^{B}(\bm{X}_{N_F}, \bm{x})\bm{k}^{B}(\bm{x},\bm{X}_{N_F}) d\bm{x} & \bm{0}_{N_F \times N_M}\\
\bm{0}_{N_M, \times N_F} & \bm{0}_{N_M \times N_M}
\end{bmatrix}
\notag
\end{equation}

Evaluation of all entries in \(\bm{W}^{B,B}\) is only related to \(\bm{X}_N\) space. For
the Gaussian kernel, the integral required for evaluation is equivalent
to \eqref{eq:wmmint}. In the search to find the additional data point
\(\tilde{\bm{x}}\) which minimizes KOH-IMSPE,
\(\frac{\partial \bm{W}^{B,B}}{\partial \tilde{\bm{x}}} = \bm{0}\). If acquiring field data while utilizing the information obtained from a computer simulator, derivatives would be equivalent to those for \(\bm{W}^{M,M}(\bm{X}_N)\).

\hypertarget{blockmatrixA}{%
\subsection{Block Matrix Inversion}\label{blockmatrixA}}

Let:
\begin{equation}
\bm{\Sigma}^{M,B}_{N_F + N_M + 1} = 
\begin{bmatrix}
\bm{\Sigma}^{M,B}_{N_F + N_M} & \nu_M \tilde{\bm{k}}\\[5pt]
\nu_M \tilde{\bm{k}}^{\top} & \nu_M \bm{k}([\tilde{\bm{x}}, \tilde{\bm{u}}])
\end{bmatrix}
\notag
\end{equation}
where
\begin{equation}
\tilde{\bm{k}} =
\begin{bmatrix}
\bm{k}([\bm{X}_{N_F}, \hat{\bm{U}}], [\tilde{\bm{x}}, \tilde{\bm{u}}])\\[5pt]
\bm{k}([\bm{X}_{N_M}, \bm{U}_{N_M}], [\tilde{\bm{x}}, \tilde{\bm{u}}])
\end{bmatrix}
\end{equation}
We can then find \([\bm{\Sigma}_{N_F + N_M + 1}^{M,B}]^{-1}\) as such:
\begin{equation}
[\bm{\Sigma}_{N_F + N_M + 1}^{M,B}]^{-1} =
\begin{bmatrix}
\kern2pt [\bm{\Sigma}^{M,B}_{N_F + N_M}]^{-1} + \frac{1}{b}\nu_M^2[\bm{\Sigma}^{M,B}_{N_F + N_M}]^{-1}\tilde{\bm{k}}\tilde{\bm{k}}^{\top}[\bm{\Sigma}^{M,B}_{N_F + N_M}]^{-1} & -\frac{1}{b}\nu_M[\bm{\Sigma}^{M,B}_{N_F + N_M}]^{-1}\tilde{\bm{k}}\kern2pt\\[5pt]
\kern2pt -\frac{1}{b}\nu_M\tilde{\bm{k}}^{\top}[\bm{\Sigma}^{M,B}_{N_F + N_M}]^{-1} & \frac{1}{b}\kern2pt
\end{bmatrix},
\label{eq:bmatrixinv}
\end{equation}
where \(b = \nu_M \bm{k}([\tilde{\bm{x}}, \tilde{\bm{u}}]) - \nu_M^2 \tilde{\bm{k}}^{\top}[\bm{\Sigma}^{M,B}_{N_F + N_M}]^{-1}\tilde{\bm{k}}\).

When minimizing KOH-IMSPE, it is necessary to differentiate \([\bm{\Sigma}^{M,B}_{N_F + N_M + 1}]^{-1}\), and therefore the differential must be pushed through the block matrix inverse form \eqref{eq:bmatrixinv} in order to reduce the expense of gradient evaluations. Differentiating each of the blocks individually provides the following results via the use of the chain rule. It is important to note, that when \(\frac{\partial[\bm{\Sigma}^{M,B}_{N_F + N_M + 1}]^{-1}}{[\tilde{\bm{x}}, \tilde{\bm{u}}]_l}\) can be found in closed form, the identity \(\frac{\partial \bm{U}(x)^{-1}}{\partial x} = -\bm{U}(x)^{-1} \frac{\partial \bm{U}(x)}{\partial x}\bm{U}(x)^{-1}\) is not necessary in \eqref{eq:imspediff}.
\begin{equation}
\frac{\partial b^{-1}}{\partial [\tilde{\bm{x}}, \tilde{u}]_l} = b^{-2}\nu_M^2\left(\tilde{\bm{k}}^{\top}[\bm{\Sigma}^{M,B}_{N_F + N_M}]^{-1}\frac{\partial \tilde{\bm{k}}}{\partial [\tilde{\bm{x}}, \tilde{\bm{u}}]_l} + \frac{\partial \tilde{\bm{k}}^{\top}}{\partial [\tilde{\bm{x}}, \tilde{\bm{u}}]_l}[\bm{\Sigma}^{M,B}_{N_F + N_M}]^{-1} \tilde{\bm{k}}\right) 
\notag
\end{equation}
\begin{equation}
\begin{split}
\frac{\partial \left(b^{-1}\nu_M^2[\bm{\Sigma}^{M,B}_{N_F + N_M}]^{-1}\tilde{\bm{k}}\tilde{\bm{k}}^{\top}[\bm{\Sigma}^{M,B}_{N_F + N_M}]^{-1}\right)}{\partial [\tilde{\bm{x}}, \tilde{\bm{u}}]_l} &= b^{-1}\nu_M^2[\bm{\Sigma}^{M,B}_{N_F + N_M}]^{-1}\tilde{\bm{k}}\frac{\partial \tilde{\bm{k}}^{\top}}{\partial[\tilde{\bm{x}}, \tilde{\bm{u}}]_l}[\bm{\Sigma}^{M,B}_{N_F + N_M}]^{-1} \\
& \qquad + \frac{\partial b^{-1}}{\partial [\tilde{\bm{x}}, \tilde{\bm{u}}]_l} \nu_M^2[\bm{\Sigma}^{M,B}_{N_F + N_M}]^{-1}\tilde{\bm{k}}\tilde{\bm{k}}^{\top}[\bm{\Sigma}^{M,B}_{N_F + N_M}]^{-1} \\
& \qquad + b^{-1}\nu_M^2[\bm{\Sigma}^{M,B}_{N_F + N_M}]^{-1}\frac{\partial \tilde{\bm{k}}}{\partial [\tilde{\bm{x}}, \tilde{\bm{u}}]_l}\tilde{\bm{k}}^{\top}[\bm{\Sigma}^{M,B}_{N_F + N_M}]^{-1}
\end{split}
\notag
\end{equation}
\begin{equation}
\frac{\partial \left(b^{-1}\nu_M[\bm{\Sigma}^{M,B}_{N_F + N_M}]^{-1}\tilde{\bm{k}}\right)}{\partial[\tilde{\bm{x}}, \tilde{\bm{u}}]_l} = \frac{\partial b^{-1}}{\partial [\tilde{\bm{x}}, \tilde{\bm{u}}]_l} \nu_M[\bm{\Sigma}^{M,B}_{N_F + N_M}]^{-1}\tilde{\bm{k}} + b^{-1}\nu_M[\bm{\Sigma}^{M,B}_{N_F + N_M}]^{-1}\frac{\partial \tilde{\bm{k}}}{\partial [\tilde{\bm{x}}, \tilde{\bm{u}}]_l}
\notag
\end{equation}
\begin{equation}
\frac{\partial \left(b^{-1}\nu_M\tilde{\bm{k}}^{\top}[\bm{\Sigma}^{M,B}_{N_F + N_M}]^{-1}\right)}{\partial [\tilde{\bm{x}}, \tilde{\bm{u}}]_l} = \left(\frac{\partial \left(b^{-1}\nu_M[\bm{\Sigma}^{M,B}_{N_F + N_M}]^{-1}\tilde{\bm{k}}\right)}{\partial[\tilde{\bm{x}}, \tilde{\bm{u}}]_l}\right)^{\top}
\notag
\end{equation}

\hypertarget{sxappendix}{%
\section{Solvent Extraction Modeling}\label{sxappendix}}

\hypertarget{field-data-experimental-methods}{%
\subsection{Field data experimental methods}\label{field-data-experimental-methods}}

To generate field data, shake tests were conducted, where an aqueous solution containing various ions of interest were mixed with a liquid organic phase. During this time, ions in each phase react and are transported between phases based on the solubility of the reactants at equilibrium. The phases then disengage over time (think mixing oil and water), and a sample of the aqueous solution can be obtained for analysis via Inductively coupled plasma mass spectrometry (ICP-MS) to obtain an assay of the concentration of elements remaining in the aqueous solution.

To obtain a varying data set the ratios of the volume of organic to aqueous liquids (O/A) were varied, along with the equilibrium pH. The O/A ratios tested are 0.1 and 1. To vary the aqueous equilibrium pH after reacting with an organophosphorus acid, additions of 5\% and 50\% (m/m) sodium hydroxide (NaOH) solutions were used. The independent variables related to NaOH addition then become mols of NaOH and volume of NaOH solution added to the mixture. For modeling purposes the equilibrium pH is later treated as a variable dependent on the NaOH additions. Target equilibrium pH values used are 1.5, 2.0, 2.5, 3.0, and 3.5.

Constants between experiments include the initial elemental concentrations and pH of the aqueous solution, the initial content of the organic phase, and temperature. A single batch of the aqueous solution was mixed by dissolving a mixture of rare earth oxides in hydrochloric acid. The same aqueous solution was used for all experiments. The pH of the solution used as a feed to the process averaged to be 1.99. Multiple samples were drawn from the solution and pH was checked over time to ensure consistency. Similarly, for the organic phase the organic dilutant Elixore 205 was mixed with 2-ethylhexyl phosphonic acid mono-2-ethylhexyl ester (EHEHPA) such that the molarity of EHEHPA in the fresh organic phase is 1.053 mol/L.

\hypertarget{simulation}{%
\subsection{Simulation}\label{simulation}}

A simulator for the chemical reaction was written in the \textsf{R} programming language, where a set of differential equations was solved using a Runge-Kutta routine. The set of differential equations specified are assumed to be governed by the law of mass action \citep{espenson1995chemical}, shown as Equation set \eqref{eq:law-ma}. The law of mass action provides a means for writing a set of differential equations given the concentrations \(R_i\) and stochiometric coefficients \(r_i\) of the reactants, and reaction kinetic constants. \begin{equation}
\begin{split}
r_1R_1 + r_2R_2 + \dots + r_mR_m \overset{k}{\rightarrow} p_1P_1 + p_2P_2 + \dots + p_nP_n \\
r = -\frac{1}{r_i}\frac{d\lbrack R_i \rbrack}{dt} = \frac{1}{p_j}\frac{d\lbrack P_j\rbrack}{d_t} = k \prod_{l = 1}^m \lbrack R_l \rbrack^{r_l}
\end{split}
\label{eq:law-ma}
\end{equation}
Although data for a variety of elements is available, we choose to simplify modeling in testing the utility of KOH-IMSPE by only focusing on sodium (Na) and Lanthanum (La) concentrations. The reactions specified for simulation are based on the stated reaction of a rare earth element with an organophosphorus acid in \citet{gupta1992extractive}. The disassociation of EHEHPA when reacting with Na\(^{+1}\) is assumed to be similar to when reacting with rare earth elements. Reversible reactions used are shown as Equation set \eqref{eq:sxreactions}, where the subscript aq and org denote presence in the aqueous or organic phase, \(\mathrm{H_2A_2}\) is the organophosporus acid EHEHPA, \(\mathrm{HA_2}^-\) is the conjugate base, and \(\mathrm{H}^+\) is a hydrogen ion.\begin{equation}
\begin{split}
[\mathrm{La}^{3+}]_{\mathrm{aq}} + 3\lbrack \mathrm{H}_2\mathrm{A}_2\rbrack_{\mathrm{org}} &\overset{k^\mathrm{La}_{-}}{\underset{k^\mathrm{La}_{+}}{\rightleftharpoons}} \lbrack \mathrm{La}(\mathrm{HA}_2)_3\rbrack_{\mathrm{org}} + 3[\mathrm{H}^+]_\mathrm{aq}\\
[\mathrm{Na}^+]_\mathrm{aq} + \lbrack \mathrm{H}_2\mathrm{A}_2\rbrack_{\mathrm{org}} &\overset{k^\mathrm{Na}_{-}}{\underset{k^\mathrm{Na}_{+}}{\rightleftharpoons}} \lbrack \mathrm{Na}(\mathrm{HA}_2)\rbrack_{\mathrm{org}} + [\mathrm{H}^+]_\mathrm{aq}\\
\end{split}
\label{eq:sxreactions}
\end{equation}
Constraints can then be specified to reduce the number of differential equations which need to be solved and ensure the conservation of mass for the chemical reaction. We assume that initial concentrations of \(\lbrack \mathrm{Na(HA_2)}\rbrack_0\) and \(\lbrack \mathrm{La(HA_2)_3}\rbrack_0\) in the organic phase are zero. The set of constraints, given initial concentrations at time 0 and concentrations at time \(T\), is shown in Equation Set \eqref{eq:zeroconst}. For compactness, we remove the \(_\mathrm{aq}\) and \(_\mathrm{org}\) subscripts in all following equations, but chemical formulas remain consistent.
\begin{equation}
\begin{split}
[ \mathrm{La}^{3+}]_0 &= [ \mathrm{La}^{3+}]_T + [\mathrm{La(HA_2)_3}]_T\\
[\mathrm{Na}^+]_0 &= [\mathrm{Na}^+]_T + [\mathrm{Na(HA_2)}]_T\\
[\mathrm{H_2A_2}]_0 &= 3[\mathrm{La(HA_2)_3}]_T + [\mathrm{Na(HA_2)}]_T + [\mathrm{H_2A_2}]_T\\
[\mathrm{H}^+]_0 + [\mathrm{H_2A_2}]_0 &= [\mathrm{H}^+]_T + [\mathrm{H_2A_2}]_T
\end{split}
\label{eq:zeroconst}
\end{equation}
\eqref{eq:zeroconst} can be rearranged to solve for \([\mathrm{La(HA_2)_3}]_T, \ [\mathrm{Na(HA_2)}]_T, \ [\mathrm{H_2A_2}]_T, \ \mathrm{and} \ [\mathrm{H}^+]_T\). Importantly, the concentrations related to the organic phase are difficult or impossible to measure.
\begin{equation}
\begin{split}
[\mathrm{La(HA_2)_3}]_T &= [\mathrm{La}^{3+}]_0 - [ \mathrm{La}^{3+}]_T\\
[\mathrm{Na(HA_2)}]_T &= [\mathrm{Na}^+]_0 - [\mathrm{Na}^+]_T\\
[\mathrm{H_2A_2}]_T &= [\mathrm{H_2A_2}]_0 -  3[\mathrm{La(HA_2)_3}]_T - [\mathrm{Na(HA_2)}]_T\\
[\mathrm{H}^+]_T  &= [\mathrm{H}^+]_0 + [\mathrm{H_2A_2}]_0 - [\mathrm{H_2A_2}]_T
\end{split}
\label{eq:diffeqconst}
\end{equation}
The constraints stipulated in \eqref{eq:diffeqconst} reduce the number of differential equations which must be solved numerically to two. These equations are shown in \eqref{eq:diffeq}.
\begin{equation}
\begin{split}
\frac{d[\mathrm{La}^{3+}]}{dt} &= k_{+}^{\mathrm{La}}[\mathrm{La(HA_2)}_3]_T[\mathrm{H}^+]_T^3 - k_{-}^{\mathrm{La}}[\mathrm{La}^{3+}]_T[\mathrm{H_2A_2}]_T^3\\
\frac{d[\mathrm{Na}^{+}]}{dt} &= k_{+}^{\mathrm{Na}}[\mathrm{Na(HA_2)}]_T[\mathrm{H}^+]_T - k_{-}^{\mathrm{Na}}[\mathrm{Na}^{+}]_T[\mathrm{H_2A_2}]_T
\end{split}
\label{eq:diffeq}
\end{equation}
This set of differential equations was solved using a Runge-Kutta routine where the constraints in \eqref{eq:diffeqconst} were substituted in whenever possible. A total time of 20 was used with a step size of \(5 \times 10^{-5}\). Bounds for the chemical kinetic constants were chosen to be \(\left[10^{-3}, 10^3\right]\).

In application, each shake test is simulated as a perfectly mixed solution. Therefore, although initial concentrations in the aqueous and organic phases are constant, \([\mathrm{La}^{3+}]_0\) and \([\mathrm{H_2A_2}]_0\) will vary with a change in O/A, because original concentrations are normalized to the total volume of the organic phase, aqueous phase, and volume of NaOH solution added. The inputs to the simulator are mols of NaOH added to the system, volume of NaOH added to the system, and O/A. Initial pH is taken to be 1.99. \([\mathrm{H}^+]_0\) is calculated by taking \(10^{-1.99}\), multiplying by the aqueous volume to find mols, subtracting the mols of NaOH added, then normalizing to the total volume of the mixture. If this calculation provides a negative value then \([\mathrm{H}^+]_0\) is set equal to \(10^{-14}/(\mathrm{mols \ of \ NaOH} - 10^{-1.99} \times \mathrm{aqueous \ volume})\) normalized by the total mixture volume. Outputs of the simulator are the natural logarithm of \([\mathrm{La}]\) and \([\mathrm{Na}]\) normalized to the total aqueous volume.

\hypertarget{additional-experiments}{%
\section{Additional Experiments}\label{additional-experiments}}

Here we provide several additional experiments that provide additional
variations beyond the ones in the main body of the paper.

\hypertarget{additional-methods-for-sinusoid}{%
\subsection{Additional Methods for Sinusoid}\label{additional-methods-for-sinusoid}}

KOH-IMSPE, LHS, M-IMSPE, and M-IMSPE\(\; \mid \tilde{\mathbf{u}} =
\hat{\mathbf{u}}\) designs were compared in a MC experiment described
in Section \ref{sinexample}. In addition, uniform random, MaxPro, and an
M-IMSPE design where the acquisition in \(X\) space must be one of the
observation locations for the field data (M-IMSPE\(\;\mid \tilde{x} \in
\bm{X}_{N_F}\)) were tested within the same experiment.  For each MC run, the
field data and initial simulator design was the same for each of the
comparators.  Additionally, KOH was fit in the same manner, sharing the same
prior parameters, for each design method. Details are provided in the main
body text.

\begin{figure}[ht!]
{\centering
\includegraphics{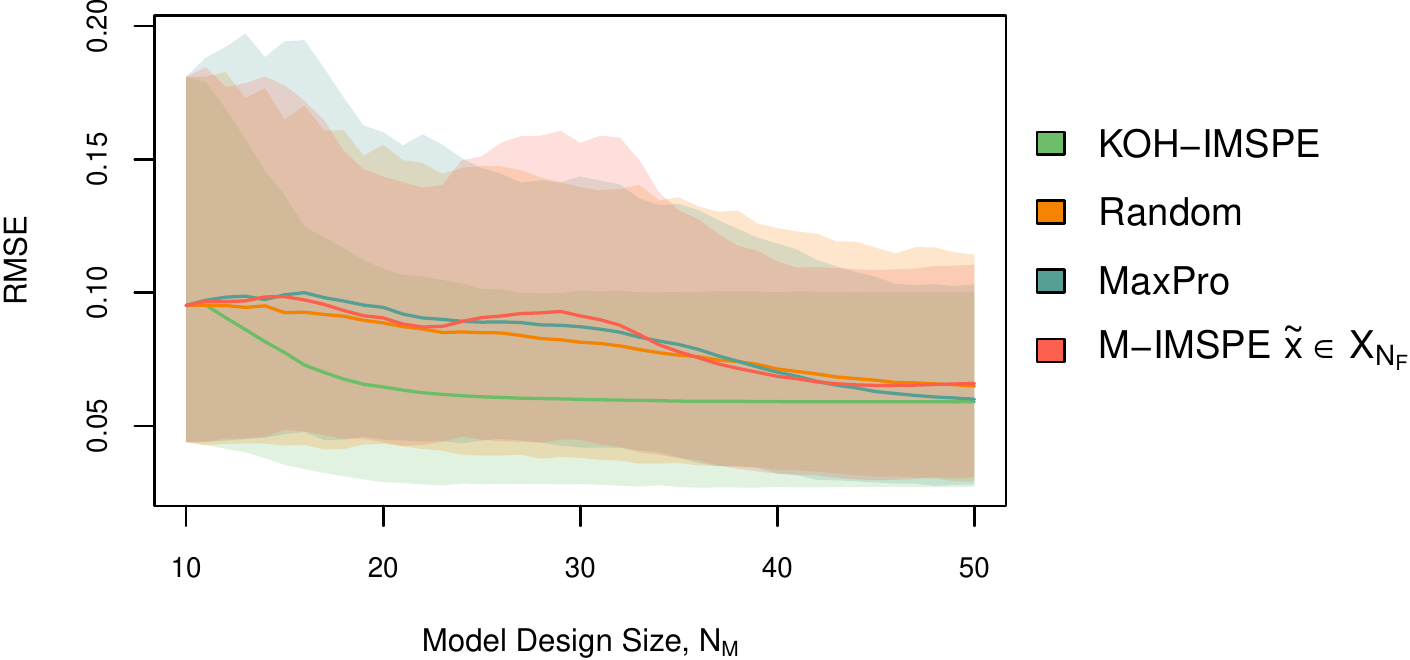} 
}\caption{Left: mean
RMSE and 90\% quantiles for sinusoid data generating mechanism using
KOH-IMSPE, random uniform, MaxPro, and M-IMSPE \(\; \mid\tilde{x} \in
\bm{X}_{N_F}\).}\label{fig:sinusoidsrmseappendix}
\end{figure}

Results for comparators not included in the main body text, in comparison to
the KOH-IMSPE results, are shown in Figure \ref{fig:sinusoidsrmseappendix}.
We can see that KOH-IMSPE on average performs better than these additional
comparators.  M-IMSPE\(|\tilde{x} \in \bm{X}_{N_F}\) struggles to improve upon
the results from the random uniform and MaxPro designs.  M-IMSPE\(|\tilde{u} =
\hat{u}\), shown in Figure \ref{fig:sinusoidsrmse}, appears to be the runner
up to KOH-IMSPE for this specific problem.

\hypertarget{additional-designs-for-gohbastos-problem}{%
\subsection{Additional Designs for Goh/Bastos Problem}\label{additional-designs-for-gohbastos-problem}}

Just like the sinusoid problem (Sections \ref{sinexample} and
\ref{additional-methods-for-sinusoid}), additional comparators not shown in
the main text body were gathered in the MC experiment described in Section
\ref{gohbastos-problem} were tested and the results are shown herein.  The
additional design methods used are random uniform, MaxPro, and
M-IMSPE\(|\tilde{\bm{x}} \in \bm{X}_{N_F}\).  All design methods shared the
same initial simulator design and field data within a single Monte Carlo
iteration.  Prior distributions used for fitting the KOH GPs are the same as
described in Section \ref{gohbastos-problem}.

\begin{figure}[ht!]
{\centering
\includegraphics{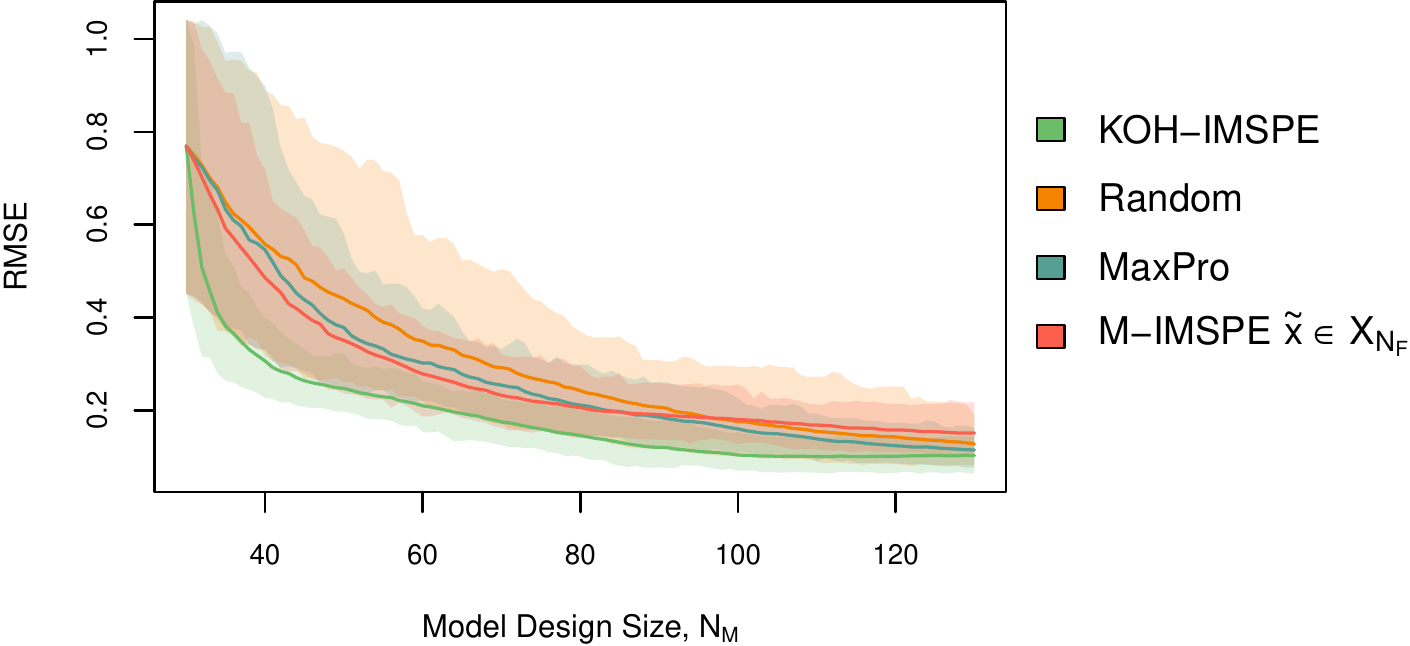} 
}\caption{Left: mean RMSE and 90\% quantiles for data generating mechanism from \citet{gramacy2020surrogates} (Ch 8, Ex 2) using KOH-IMSPE, random uniform, MaxPro, and M-IMSPE \(\mid \tilde{\bm{x}} \in \bm{X}_{N_F}\) designs.}\label{fig:surrogatesrmseappendix}
\end{figure}

The results for RMSE values for increasing \(N_M\) for the additional comparators, in addition KOH-IMSPE, are shown in Figure \ref{fig:surrogatesrmseappendix}.  KOH-IMSPE performs better than these additional comparators on average and with less variance, meaning KOH-IMSPE shows the best performance for this problem out of all the methodologies tested.  M-IMSPE, the two M-IMSPE variants, and MaxPro are all competitive with LHS and random uniform performing the most poorly.

\hypertarget{confounding}{%
\subsection{Confounding}\label{confounding}}

In an effort to examine the effect of varying the magnitude of the true bias function, a slightly modified version of the experiment in Section \ref{sinexample} was performed. In this experiment, the data generating mechanism takes the form of \(Y^F(x) = y^M(x, u^{\star}) + \alpha b(x) + \epsilon\) for \(\alpha \in \{0,1,2\}\). RMSE and estimation accuracy of \(u^{\star}\) for KOH-IMSPE relative to M-IMSPE was compared for the case of when there was no bias and a greater bias than what was previously examined.  At each MC iteration, an initial computer model design was collected as a 10 point subset of a 100 LHS. Field data was collected on a 5 point grid with two replicates at each location. When generating field data, the deterministic component of the bias function was evaluated and multiplied by the relevant \(\alpha\). 
\begin{figure}[ht!]
\includegraphics{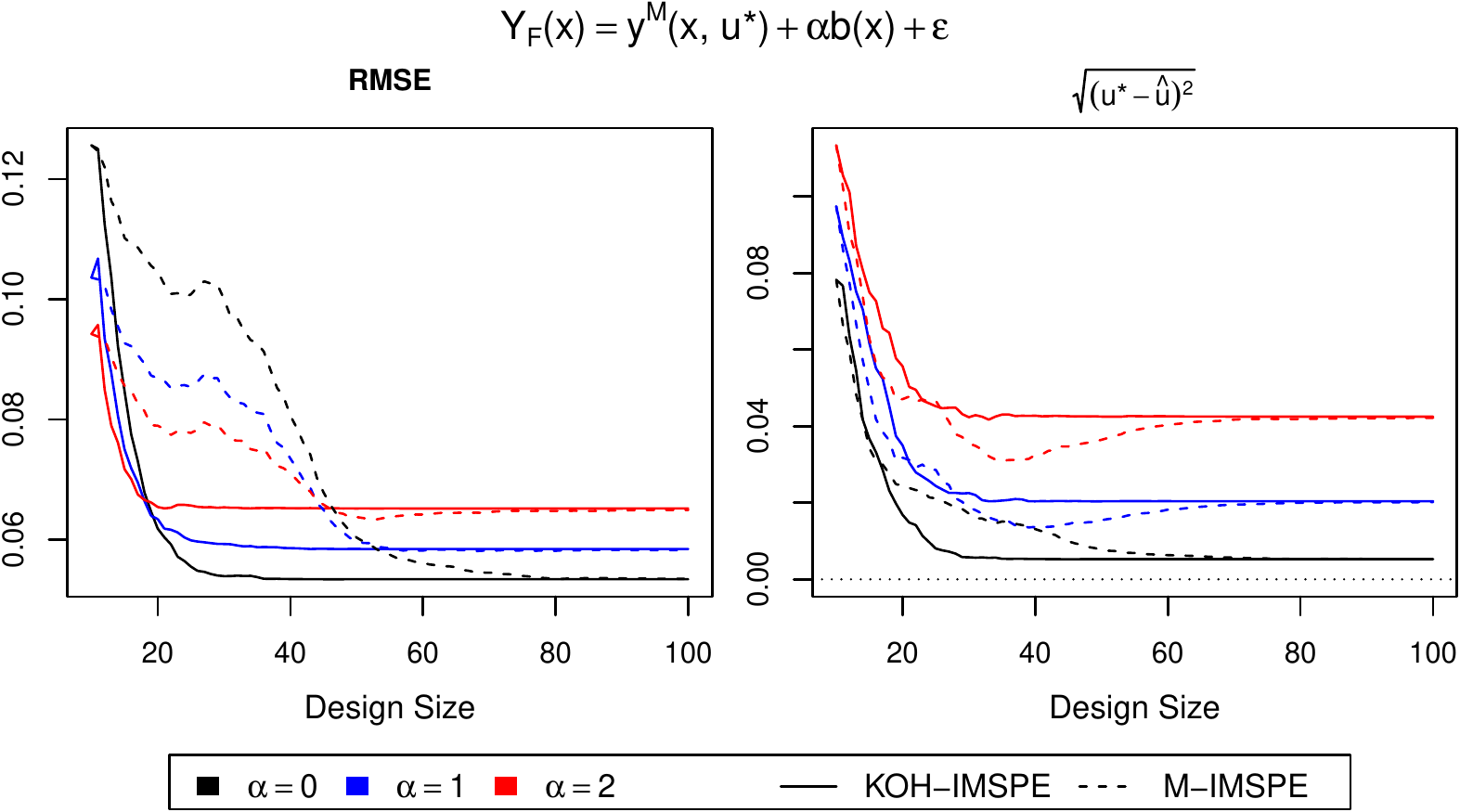} \caption{Performance measures from active learning designs where the magnitude of the bias within the data generating mechanism (\(\alpha\)) is chosen to be one of three settings such that the results can be compared.  Left:  Mean RMSE for increasing \(N_M\) using both KOH-IMSPE and standard M-IMSPE designs.  Right:  Mean distance between \(\hat{u}\) and \(u^{\star}\) for increasing \(N_M\) using both KOH-IMSPE and standard M-IMSPE designs.}\label{fig:sinuhatconfound}
\end{figure}
To reduce variability, the observed normally distributed error, \(\epsilon\), was the same for all design methods and levels of \(\alpha\) used within an individual MC iteration. After each point was added, all GPs were refit and a new value of \(\hat{u}\) was estimated. All GPs were fit and \(\hat{u}\) was estimated using a modularized KOH setup, even in the case for \(\alpha = 0\). Therefore, the behavior of KOH-IMSPE was tested in the case where the true bias between the computer model and the field data is zero, but the statistician assumes there is bias. In an effort to reduce the influence of the prior on \(\hat{u}\),  \(p(\hat{u}) = \mathrm{Beta}(1,1)\). The sequential design continued until 100 computer model points were obtained, with RMSE calculated on a 100 point noiseless hold out set after each acquisition. The methodology was repeated for 250 MC iterations.  

The mean, taken over all MC iterations, for RMSE and distance between \(\hat{u}\) and \(u^{\star}\) were plotted for increasing \(N_M\), as shown in Figure \ref{fig:sinuhatconfound}.  Figure \ref{fig:sinuhatconfound} first illustrates that KOH-IMSPE converges to the \textit{optimal} RMSE more quickly than M-IMSPE, independent of the magnitude of the bias tested.  For both design methods, a larger bias decreases the possible improvements attainable for a KOH model.  The right panel shows KOH-IMSPE converging to an \textit{optimal} \(\hat{u}\) more quickly than M-IMSPE for varying bias magnitudes.  Neither design methods address the identifiability issues caused by KOH modeling, clearly identifiability is an issue with KOH itself.  Interestingly, for \(\alpha = 0\) the average \(\hat{u}\) never quite reaches \(u^{\star}\), likely due to bias function competing with the nugget to explain the observed data within the model mis-specification.  We believe the dips below the \textit{optimal} RMSE and \(\hat{u}\) estimates for M-IMSPE are artifacts of the particular problem in conjunction with MC variance in the results.

\end{document}